\documentclass[aps,prd,onecolumn,groupedaddress,showpacs,nofootinbib,amssymb]{revtex4-2}
\usepackage[dvips]{graphicx}
\usepackage{amssymb}
\usepackage{amsmath}
\usepackage{graphicx,,color}
\usepackage{amsfonts}
\usepackage{bm}
\usepackage{cancel}
\usepackage{comment}
\usepackage{braket}

\newcommand\be{\begin{equation}}
\newcommand\ee{\end{equation}}

\newcommand\e{\mathrm{e}}
\DeclareMathOperator{\cl}{C.L}

\allowdisplaybreaks[4]

\begin{document}

\tolerance=5000

\title{Inflationary Dynamics and Swampland Criteria for Modified Gauss-Bonnet Gravity Compatible with
GW170817}
\author{S.D. Odintsov,$^{1,2}$}
\email{odintsov@ice.cat}
\author{V.K. Oikonomou$^{3}$}\,
\email{voikonomou@gapps.auth.gr;voikonomou@auth.gr;v.k.oikonomou1979@gmail.com}
\author{F.P. Fronimos$^{3}$}\,
\email{fotisfronimos@gmail.com}
\affiliation{$^{1)}$ ICREA, Passeig Luis Companys, 23, 08010 Barcelona, Spain\\
$^{2)}$ Institute of Space Sciences (ICE, CSIC) C. Can Magrans
s/n, 08193 Barcelona, Spain\\
$^{3)}$Department of Physics, Aristotle University of
Thessaloniki, Thessaloniki 54124, Greece}

\tolerance=5000

\begin{abstract}
In this article we present an alternative formalism for the
inflationary phenomenology of rescaled Einstein-Gauss-Bonnet
models which are in agreement with the GW170817 event. By
constraining the propagation velocity of primordial tensor
perturbations, an approximate form for the time derivative of the
scalar field coupled to the Gauss-Bonnet density is extracted. In
turn, the overall degrees of freedom decrease and similar to the
case of the canonical scalar field, only one scalar function needs
to be designated, while the other is extracted from the continuity
equation of the scalar field. We showcase explicitly that the
slow-roll indices can be written in a closed form as functions of
three dimensionless parameters, namely
$x=\frac{1}{2\alpha}\bigg(\frac{\kappa\xi'}{\xi''}\bigg)^2$,
$\beta=8H^2\xi''$ and $\gamma=\frac{\xi'\xi'''}{\xi''^2}$ and in
turn, we prove that the Einstein-Gauss-Bonnet model can in fact
produce a blue-tilted tensor spectral index if the condition
$\beta\geq1$ is satisfied, which is possible only for
Einstein-Gauss-Bonnet models with $\xi''(\phi_k)>0$. Afterwards, a
brief comment on the running of the spectral indices is made where
it is shown that $a_{\mathcal{S}}(k_*)$ and $a_{\mathcal{T}}(k_*)$
in the constrained case are approximately of the order
$\mathcal{O}(10^{-3})$, if not smaller. Last but not least, we
examine the conditions under which the Swampland criteria are
satisfied. We connect the tracking condition related to scalar
field theories with the present models, and we highlight the
important feature of the models we propose that the tracking
condition can be satisfied only if the Swampland criteria are
simultaneously satisfied, however the cases with $\xi\sim1/V$ and
$\xi\sim V$ are excluded, as they cannot describe the inflationary
era properly. Also, even though the Swampland criteria can be in
agreement with a blue-tilted tensor spectral index, we prove that
there exists no model that respects the tracking condition while
at the same time it manifests a blue-tilted tensor spectrum. We
also discuss the preheating and reheating era implications and
also issues related to the energy spectrum of primordial
gravitational waves.
\end{abstract}


\maketitle

\section{Introduction}

Currently, the maim focus in theoretical particle physics and
cosmology is on observations related to the inflationary era.
Inflation itself
\cite{inflation1,inflation2,inflation3,inflation4} is a consistent
theoretical proposal towards describing the early Universe, which
solves a lot of shortcomings of the standard hot Big Bang
cosmology, however, to date it is not observationally verified.
There are two smoking gun observations that will verify the
occurrence of the inflationary era, the direct detection of the
$B$-modes in the Cosmic Microwave Background (CMB) radiation
polarization pattern, or a direct detection of primordial tensor
modes by some of the future gravitational wave experiments
\cite{Baker:2019nia,Smith:2019wny,Crowder:2005nr,Smith:2016jqs,Seto:2001qf,Kawamura:2020pcg}.
The stage four CMB experiments
\cite{CMB-S4:2016ple,SimonsObservatory:2019qwx} will further seek
for $B$-modes and will further constrain inflation. Thus the urge
for finding viable inflationary models that can predict detectable
primordial gravitational waves is compelling, since in the case of
finding a signal in the future experiments, this must be explained
theoretically. One class of candidate theories that yields a
detectable gravitational wave signal is the Einstein-Gauss-Bonnet
theory
\cite{Hwang:2005hb,Nojiri:2006je,Cognola:2006sp,Nojiri:2005vv,Nojiri:2005jg,Satoh:2007gn,Bamba:2014zoa,Yi:2018gse,Guo:2009uk,Guo:2010jr,Jiang:2013gza,Kanti:2015pda,vandeBruck:2017voa,Kanti:1998jd,Pozdeeva:2020apf,Vernov:2021hxo,Pozdeeva:2021iwc,Koh:2014bka,Bayarsaikhan:2020jww,Tumurtushaa:2018lnv,Fomin:2020hfh,DeLaurentis:2015fea,Chervon:2019sey,Nozari:2017rta,Odintsov:2018zhw,Kawai:1998ab,Yi:2018dhl,vandeBruck:2016xvt,Kleihaus:2019rbg,Bakopoulos:2019tvc,Maeda:2011zn,Bakopoulos:2020dfg,Ai:2020peo,Oikonomou:2020oil,Odintsov:2020xji,Oikonomou:2020sij,Odintsov:2020zkl,Odintsov:2020mkz,Venikoudis:2021irr,Kong:2021qiu,Easther:1996yd,Antoniadis:1993jc,Antoniadis:1990uu,Kanti:1995vq,Kanti:1997br,Easson:2020mpq,Rashidi:2020wwg},
which in most cases results in a blue-tilted tensor spectrum
\cite{Oikonomou:2022xoq}. Such theories though are plagued with
the issue of having a gravitational wave speed different from that
of light's in the vacuum, which after the GW170817 event
\cite{TheLIGOScientific:2017qsa,Monitor:2017mdv,GBM:2017lvd}, are
considered less appealing, if not an incorrect description of
nature. For a complete list of theories that were excluded after
the GW170817 event, see for example
\cite{Ezquiaga:2017ekz,Baker:2017hug,Creminelli:2017sry,Sakstein:2017xjx}.
However, such theories can be remedied if the propagation speed of
the inflationary tensor modes are equal to unity in natural units
(equal to that of light's in vacuum and in natural units $c=1$),
and this can be achieved if their non-trivial Gauss-Bonnet
coupling function, usually denoted as $\xi (\phi)$, is constrained
\cite{Odintsov:2020sqy,Oikonomou:2021kql,Oikonomou:2022ksx}. In
this article we shall present an alternative and refined formalism
for studying the inflationary phenomenology of
Einstein-Gauss-Bonnet theories which have the Ricci scalar term
rescaled by a coupling constant $\alpha$. The agreement of the
theory with the GW170817 will be a prerequisite for these
theories, thus the coupling function and the scalar potential are
interrelated and should not be freely chosen in an independent
way. Hence, the total degrees of freedom of the model decrease
and, similar to the case of the canonical scalar field, only the
scalar potential needs to be specified, while the other is
extracted from the continuity equation of the scalar field. We
demonstrate explicitly that the slow-roll indices can be written
in a closed form as functions of three dimensionless parameters,
namely
$x=\frac{1}{2\alpha}\bigg(\frac{\kappa\xi'}{\xi''}\bigg)^2$,
$\beta=8H^2\xi''$ and $\gamma=\frac{\xi'\xi'''}{\xi''^2}$ and in
turn, we prove that the Einstein-Gauss-Bonnet model can in fact
produce a blue-tilted tensor spectral index if the condition
$\beta\geq1$ is satisfied, which is possible only for
Einstein-Gauss-Bonnet models with $\xi''(\phi_k)>0$. The running
of the spectral indices is also considered, in which case it is
shown that $a_{\mathcal{S}}(k_*)$ and $a_{\mathcal{T}}(k_*)$ in
the constrained case are approximately of the order
$\mathcal{O}(10^{-3})$, if not smaller. Finally, we examine the
conditions under which the Swampland criteria are satisfied
\cite{Vafa:2005ui,Ooguri:2006in,Palti:2020qlc,Brandenberger:2020oav,Blumenhagen:2019vgj,Wang:2019eym,Benetti:2019smr,Palti:2019pca,Cai:2018ebs,Akrami:2018ylq,Mizuno:2019pcm,Aragam:2019khr,Brahma:2019mdd,Mukhopadhyay:2019cai,Brahma:2019kch,Haque:2019prw,Heckman:2019dsj,Acharya:2018deu,Elizalde:2018dvw,Cheong:2018udx,Heckman:2018mxl,Kinney:2018nny,Garg:2018reu,Lin:2018rnx,Park:2018fuj,Olguin-Tejo:2018pfq,Fukuda:2018haz,Wang:2018kly,Ooguri:2018wrx,Matsui:2018xwa,Obied:2018sgi,Agrawal:2018own,Murayama:2018lie,Marsh:2018kub,Storm:2020gtv,Trivedi:2020wxf,Sharma:2020wba,Mohammadi:2020twg,Trivedi:2020xlh,Han:2018yrk,Achucarro:2018vey,Akrami:2020zfz,Colgain:2018wgk,Colgain:2019joh,Banerjee:2020xcn,Oikonomou:2021zfl,Gitsis:2023cyw}
. We also try to establish a connection between the tracking
condition related to scalar field theories, with the present
rescaled Einstein-Gauss-Bonnet models, and we highlight the
important feature of the models we propose, which is that the
tracking condition can be satisfied only if the Swampland criteria
are simultaneously satisfied, however the cases with $\xi\sim1/V$
and $\xi\sim V$ are excluded, as they cannot describe the
inflationary era consistently. Also, even though the Swampland
criteria can be in agreement with a blue-tilted tensor spectral
index, we prove that there exists no model that respects the
tracking condition while at the same time it results to a
blue-tilted tensor spectrum. Furthermore we discuss several issues
related to the reheating era
\cite{Venikoudis:2022gfg,ElBourakadi:2021blc,ElBourakadi:2021nyb,Koh:2018qcy,Cook:2015vqa,ElBourakadi:2022lqf,Mazumdar:2013gya,Ellis:2015pla,Hasegawa:2019jsa},
and finally the amplification of the primordial gravitational
waves is considered by theories in our theoretical framework
\cite{Odintsov:2021kup,Kamionkowski:2015yta,Turner:1993vb,Boyle:2005se,Zhang:2005nw,Caprini:2018mtu,Clarke:2020bil,Smith:2005mm,Giovannini:2008tm,Liu:2015psa,Vagnozzi:2020gtf,Kamionkowski:1993fg,Giare:2020vss,Zhao:2006mm,Lasky:2015lej,Cai:2021uup,Benetti:2021uea,Lin:2021vwc,Zhang:2021vak,Pritchard:2004qp,Khoze:2022nyt,ElBourakadi:2022anr,Arapoglu:2022vbf,Giare:2022wxq,Odintsov:2022cbm,Linder:2021pek}.
The paper focuses mainly on the derivation of the duration of the
preheating and reheating eras assuming a non canonical effective
EoS which remains constant. The contribution of the Gauss-Bonnet
term under the assumption that the scalar coupling function
satisfies a specific differential equation in order to produce
massless primordial gravitons is highlighted and also tested by
using a specific model of interest. Finally the energy spectrum of
primordial gravitational waves is discussed and as shown the
amplitude can be amplified compared to the GR description, not
only due to running of the effective Planck mass, owning to the
presence of the Gauss-Bonnet term in the gravitational action that
effective changes its value, but also due to the possibility of
generating a blue spectrum once the condition $\beta>1$ is
satisfied. The latter of course applies to the high frequency
regime and is connected to the primordial era given that high
frequency modes are the first to become subhorizon.

This paper is organized as follows: In section II we present the
formalism of the rescaled Einstein-Gauss-Bonnet theories, and we
extract the inflationary indices and the observational indices in
some detail. Related issues, such as the gravitational wave speed,
the non-Gaussianities predictions and the overall phenomenological
viability of the theory at hand are considered. In section III,
the running of the spectral indices in rescaled
Einstein-Gauss-Bonnet theories is considered, while in section IV
we discuss how the Swampland criteria can be satisfied by the
rescaled Einstein-Gauss-Bonnet theories. In section V, the
reheating era is considered in detail in the present context while
in VI, the amplification of the primordial gravitational waves is
discussed in the present context. Finally, the conclusions follow
in the end of the article.

\section{Rescaled Einstein-Gauss-Bonnet Inflationary Phenomenology}

Let us commence the study by specifying the gravitational action
of the model we shall consider. For the case at hand, we assume a
rescaled version of the standard Einstein-Gauss-Bonnet model of
the form,
\begin{equation}
\centering \label{action} \mathcal{S}=\int
d^4x\sqrt{-g}\bigg(\frac{\alpha
R}{2\kappa^2}-\frac{1}{2}g^{\mu\nu}\nabla_\mu\phi\nabla_\nu\phi-V(\phi)-\xi(\phi)\mathcal{G}\bigg)\,
,
\end{equation}
for reviews on Einstein-Gauss-Bonnet gravity see
\cite{reviews1,reviews2,reviews3,reviews4}, where
$\kappa=\frac{1}{M_{Pl}}=\sqrt{8\pi G}$ is the gravitational
constant, $M_{Pl}$ being the reduced Planck mass and $\alpha$ is a
dimensionless constant parameter introduced for the sake of
generality \cite{Pozdeeva:2020apf} and is strictly positive with
the limit $\alpha\to 1$ generating the usual contribution. Such
effective field theory models may emerge for example from an
$f(R)$ gravity at early times, with the Einstein-Hilbert term in
Eq. (\ref{action}) being replaced by
\begin{equation}
\label{action1231} \centering \mathcal{S}=\int d^4x\sqrt{-g}
\frac{f(R)}{2\kappa^2}\, ,
\end{equation}
and an example of an $f(R)$ gravity which may result to a
rescaled Einstein-Hilbert term is \cite{Oikonomou:2020oex},
\begin{equation}\label{frini}
f(R)=R-\gamma  \lambda  \Lambda -\lambda  R \exp
\left(-\frac{\gamma  \Lambda }{R}\right)-\frac{\Lambda
\left(\frac{R}{m_s^2}\right)^{\delta }}{\zeta }\, .
\end{equation}
In the large curvature limit, the exponential term of Eq.
(\ref{frini}) at leading order yields,
\begin{equation}\label{expapprox}
\lambda  R \exp \left(-\frac{\gamma  \Lambda }{R}\right)\simeq
-\gamma \lambda  \Lambda -\frac{\gamma ^3 \lambda \Lambda^3}{6
R^2}+\frac{\gamma ^2 \lambda  \Lambda ^2}{2 R}+\lambda  R\, ,
\end{equation}
hence, the effective action  during inflation contains terms of
the Ricci scalar as follows,
\begin{equation}\label{effectiveaction}
\mathcal{S}=\int
d^4x\sqrt{-g}\left(\frac{1}{2\kappa^2}\left(\alpha R+ \frac{\gamma
^3 \lambda \Lambda ^3}{6 R^2}-\frac{\gamma ^2 \lambda \Lambda
^2}{2 R}-\frac{\Lambda}{\zeta
}\left(\frac{R}{m_s^2}\right)^{\delta
}+\mathcal{O}(1/R^3)+...\right)-\frac{1}{2}g^{\mu\nu}\nabla_\mu\phi\nabla_\nu\phi-V(\phi)-\xi(\phi)\mathcal{G}\right)\,
,
\end{equation}
where $\alpha=1-\lambda$. Thus such rescaled form of
Einstein-Gauss-Bonnet gravity as the one in Eq. (\ref{action}) is
well motivated from phenomenological reasoning. Of course, the
form of $f(R)$ gravity given in Eq. (\ref{frini}) is just a
working example, but the same result can be obtained from other
forms of the $f(R)$ gravity. A good point to mention is that the
effect of a rescaled Ricci scalar term would affect the effective
gravitational Newton's constant and the constraints of the Big
Bang Nucleosynthesis would apply. However, the rescaled form of
the $F(R)$ gravity in the form of $F(R)\sim \alpha R$ is valid
only at early times, well before the commencing of the reheating
era and during the inflationary era, when the curvature is large.
In the post inflationary era, during the reheating and the
subsequent radiation domination era, the approximation in Eq.
(\ref{expapprox}) holds no longer true, thus during the BBN, which
is deeply in the radiation domination era, no rescaling affects
the evolution anymore. Thus the constraints on the effective
gravitational constant $G$ are not affected by the rescaled
gravity. This rescaled gravity is valid only during the
inflationary era and ceases to be valid at the end of inflation
and at the beginning of the reheating era.

In this work we shall approach the problem in an agnostic way and
use such a rescaling without specifying the underlying theory that
generates it. In addition, the existence of a canonical scalar
field is assumed with
$\frac{1}{2}g^{\mu\nu}\nabla_\mu\phi\nabla_\nu\phi$ and $V(\phi)$
being the kinetic term and the canonical potential, with the
latter remaining unspecified for the time being and furthermore, a
non minimal coupling between the scalar field and the Gauss-Bonnet
density  is assumed, with the latter being,
$\mathcal{G}=R^2-4R_{\mu\nu}R^{\mu\nu}+R^{\mu\nu\rho\sigma}R_{\mu\nu\rho\sigma}$
where $R_{\mu\nu}$ and $R_{\mu\nu\rho\sigma}$ are the Ricci and
Riemann tensor respectively. Note that similar to the scalar
potential, the Gauss-Bonnet scalar coupling function, which is
dimensionless, shall remain arbitrary, however in the following it
shall be constrained by demanding a specific time evolution so
that compatibility with the recent GW170817 event can be achieved.
It should also be stated that while the Gauss-Bonnet density
cannot be inserted in the gravitational action in a linear form
given that it vanishes as a surface term, the arbitrary scalar
coupling function assures that the Gauss-Bonnet term participates
both in the background equations and in the perturbed equations as
long as it dynamically evolves, that is $\dot\xi\neq 0$. It is
also worth mentioning that in recent developments, it was
showcased that the Gauss-Bonnet density can in fact be inserted
linearly in the action in $D$ dimensions and, by assuming a $\phi$
independent $\xi$, meaning no coupling between $\mathcal{G}$ and
the scalar field, a simple rescaling of $\xi$ as
$\xi\to\frac{\xi}{D-4}$ can in fact result in the appearance of
the Gauss-Bonnet term in the field equations when taking the limit
$D\to 4$, see Ref. \cite{Ai:2020peo,Easson:2020mpq}. An
interesting result is that in this approach, the propagation
velocity of gravitational waves is not affected in contrast to the
non-minimal model. This scenario, although quite interesting,
shall not be studied here and instead the non-minimal coupling
shall be considered.

Before continuing it is worth discussing the motivation in
considering Lagrangian densities containing both Gauss-Bonnet
terms and $f(R)$ gravity terms, since these Lagrangian densities
are somewhat complicated. The motivation comes from string theory
and quantum corrected scalar field Lagrangian densities. The most
general scalar field Lagrangian in four dimensions which contains
at most two derivatives has the following form,
\begin{equation}\label{generalscalarfieldaction}
\mathcal{S}_{\varphi}=\int
\mathrm{d}^4x\sqrt{-g}\left(\frac{1}{2}Z(\varphi)g^{\mu
\nu}\partial_{\mu}\varphi
\partial_{\nu}\varphi+\mathcal{V}(\varphi)+h(\varphi)\mathcal{R}
\right)\, .
\end{equation}
When the scalar fields are considered to be evaluated in their
vacuum configuration, the scalar field must be either minimally or
conformally coupled to gravity. In our case the scalar field is
minimally coupled, and in this case, the one loop quantum
corrected scalar field action consistent with diffeomorphism
invariance and which contains up to order four derivatives is the
following \cite{Codello:2015mba},
\begin{align}\label{quantumaction}
&\mathcal{S}_{eff}=\int
\mathrm{d}^4x\sqrt{-g}\Big{(}\Lambda_1+\Lambda_2
\mathcal{R}+\Lambda_3\mathcal{R}^2+\Lambda_4 \mathcal{R}_{\mu
\nu}\mathcal{R}^{\mu \nu}+\Lambda_5 \mathcal{R}_{\mu \nu \alpha
\beta}\mathcal{R}^{\mu \nu \alpha \beta}+\Lambda_6 \square
\mathcal{R}\\ \notag &
+\Lambda_7\mathcal{R}\square\mathcal{R}+\Lambda_8 \mathcal{R}_{\mu
\nu}\square \mathcal{R}^{\mu
\nu}+\Lambda_9\mathcal{R}^3+\mathcal{O}(\partial^8)+...\Big{)}\, ,
\end{align}
with the parameters $\Lambda_i$, $i=1,2,...,6$ being appropriate
dimensionful constants. Thus the rescaled gravity Lagrangian which
also contains higher order powers of Ricci and Riemann tensors, or
even theories with a combination of them in the form of the
Gauss-Bonnet invariant, result from quantum corrections of the
scalar field Lagrangian and thus are string theory motivated. Here
we consider only the overall simplified form of a simple rescaling
appearing in the $R$ term.

Furthermore, motivated by the fact that the metric corresponds to
a homogeneous and isotropic expanding universe with the
Friedmann-Robertson-Walker (FRW) line element,
\begin{equation}
\centering
\label{metric}
 ds^2=-dt^2+a^2(t)\delta_{ij}dx^idx^j\, ,
\end{equation}
where $a(t)$, not to be confused with the constant parameter
$\alpha$, is the scale factor, it will also be assumed that the
scalar field is homogeneous therefore it depends solely on cosmic
time $t$. In turn, assuming that $\phi=\phi(t)$ simplifies both
its kinetic term, which is now equal to
$\frac{1}{2}g^{\mu\nu}\nabla_\mu\phi\nabla_\nu\phi=-\frac{1}{2}\dot\phi^2$
where the ``dot'' as usual implies differentiation with respect to
$t$, and also its dynamical evolution which is dictated by the
continuity equation. Before we start working on the background
equations however, let us focus briefly on tensor perturbations
and in particular on their propagation velocity.

The inclusion of the Gauss-Bonnet density through the assumption
that a non-minimal coupling between the scalar field and curvature
exists, seems to have an interesting impact on tensor
perturbations. Since this model can safely be considered as a
specific subclass of Horndeski's theory \cite{Horndeski:1974wa},
given that the action (\ref{action}) results in a second order
differential equation for the scalar field, the propagation
velocity of gravitational waves if one performs a linear analysis
on tensor perturbations, can easily be derived and expressed in
the following form \cite{Hwang:2005hb},
\begin{equation}
\centering
\label{cT}
c_\mathcal{T}^2=1-\frac{Q_f}{Q_t}\, ,
\end{equation}
where $Q_f=8(\ddot\xi-H\dot\xi)$ and
$Q_t=\frac{\alpha}{\kappa^2}-8\dot\xi H$ with natural units being
used for convenience such that $c=1$. As shown, there exists a
deviation from the speed of light which is quantified by the
magnitude of the ratio $\frac{Q_f}{Q_t}$ and appears due to the
fact that the non-minimal coupling is considered. This realization
seems to be at variance with observations as the merging of two
neutron stars in the kilonova GW170817 event
\cite{TheLIGOScientific:2017qsa} made it abundantly clear that
gravitational waves propagate through spacetime with the velocity
of light as electromagnetic and gravitational radiation reached
Earth simultaneously. In order to reconcile the model at hand with
observations, hereafter the Gauss-Bonnet scalar coupling function
is assumed to satisfy the following differential equation,
\begin{equation}
\centering
\label{constraint1}
\ddot\xi=H\dot\xi\, .
\end{equation}
The same conclusion about tensor perturbations can be reached from
Ref. \cite{Fomin:2020hfh}, where linear perturbations were also
studied. This constraint is quite powerful as it decreases the
overall degrees of freedom \cite{Odintsov:2020sqy} and also
restores compatibility of the Einstein-Gauss-Bonnet model with
observations. It should also be stated that, while the GW170817
event is in fact observed in the late-era, the constraint can be
imposed regardless of the era studied and thus, in our case it is
imposed on the early era (see Ref. \cite{Odintsov:2021nim} for the
impact of the constraint on the late-time acceleration). This may
be bizarre as one may be inclined to believe that the constraint
itself is not needed in order to obtain compatibility with
observations. This is because the GW170817 event which as
mentioned before is a late-time observation, could be in agreement
with the Einstein-Gauss-Bonnet model provided that the scalar
field has reached its vacuum expectation value and does not evolve
dynamically. In turn, this implies that $\dot\xi=0=\ddot\xi$ and
thus Eq. (\ref{cT}) safely reaches the limit $c_\mathcal{T}=1$ without
implementing any constraint. While this statement is indeed valid,
the same argument cannot be used in the early era since during
inflation, the canonical scalar field evolves dynamically as it
drives inflation and thus the propagation velocity of tensor
perturbations cannot be equal to the speed of light. While there
exists no evidence that excludes the possibility of $c_\mathcal{T}<1$
primordially, the constraint (\ref{constraint1}) is implemented
simply because the inclusion of the non-minimal coupling between
the scalar field and the Gauss-Bonnet density $\mathcal{G}$ serves
as a low-energy effective string inspired model which would then
predict that primordial gravitons are massive. In addition, their
mass would vanish as soon as the scalar field reaches its vacuum
expectation value, suggesting the existence of a procedure which
is the inverse of the Higgs mechanism. In order to avoid the
appearance of primordial massive gravitons in the model at hand,
the constraint (\ref{constraint1}) is implemented. In turn,
assuming that the scalar field evolves slowly, one can show that
the time derivative of the scalar field is given by the following
expression \cite{Odintsov:2020sqy},
\begin{equation}
\centering
\label{constraint2}
\dot\phi\simeq\frac{H\xi'}{\xi''}\, ,
\end{equation}
which serves as a $\dot\phi(\phi)$ solution. As a result, the
continuity equation of the scalar field is not required in order
to extract algebraically $\dot\phi$, provided that $\ddot\phi\ll
H\dot\phi$, but it can be used in order to constrain one of the
scalar functions. Let us showcase this explicitly.

The gravitational action (\ref{action}) for the model at hand is a
functional of the metric tensor $g^{\mu\nu}$ and the scalar field,
that is $S=S[g^{\mu\nu},\phi]$. Therefore, by varying the
aforementioned action with respect to the metric and the scalar
field, the field equations and the continuity equation of the
scalar field are extracted which in this case read
\cite{Hwang:2005hb},
\begin{equation}
\centering
\label{fieldeq}
\frac{\alpha}{\kappa^2} G_{\mu\nu}=T^{(\xi)}_{\mu\nu}+\nabla_\mu\phi\nabla_\nu\phi-\bigg(\frac{1}{2}g^{\alpha\beta}\nabla_\alpha\phi\nabla_\beta\phi+V\bigg)g_{\mu\nu}\, ,
\end{equation}
where the stress-energy tensor of the string corrections
$T_{\mu\nu}^{(\xi)}=-\frac{2}{\sqrt{-g}}\frac{\delta(\sqrt{-g}\xi(\phi)\mathcal{G})}{\delta
g^{\mu\nu}}$ is given by the following expression,
\begin{widetext}
\begin{align}
\centering \label{Txi} &
T^{(\xi)}_{\mu\nu}=-2\bigg[\bigg(\frac{1}{2}\mathcal{G}g_{\mu\nu}+4R_{\mu\alpha}R^\alpha_\nu+4R^{\alpha\beta}R_{\mu\alpha\nu\beta}-2R_{\mu}^{\,\,\,\alpha\beta\gamma}R_{\nu\alpha\beta\gamma}-2RR_{\mu\nu}\bigg)\xi
\\ \notag &-4\bigg(\xi^{;\alpha\beta}R_{\mu\alpha\nu\beta}-\Box\xi
R_{\mu\nu}+2\xi_{;\alpha(\nu}R^\alpha_{\mu)}-\frac{1}{2}\xi_{,\mu;\nu}R\bigg)+2(2\xi_{;\alpha\beta}R^{\alpha\beta}-\Box\xi
R)g_{\mu\nu}\bigg]\, .
\end{align}
\end{widetext}
Similarly, the continuity equation of the scalar field which is
derived by varying (\ref{action}) with respect to the scalar field
reads,
\begin{equation}
\centering
\label{conteq}
\Box\phi-V'-T^{(\xi)}=0\, ,
\end{equation}
where $\Box=\nabla_\mu\nabla^\mu$ and $T^{(\xi)}=\xi'\mathcal{G}$.
From Eq. (\ref{fieldeq}), the temporal and the spatial components
corresponding to the Friedmann and the Raychaudhuri equations
respectively can be isolated and by recalling that the scalar
field is assumed to be homogeneous, the background equations for
the model at hand obtain the following forms
\cite{Odintsov:2020sqy},
\begin{equation}
\centering
\label{motion1}
\frac{3H^2(\alpha-8\kappa^2\ddot\xi)}{\kappa^2}=\frac{1}{2}\dot\phi^2+V\, ,
\end{equation}
\begin{equation}
\centering
\label{motion2}
-\frac{2\dot H(\alpha-8\kappa^2\ddot\xi)}{\kappa^2}=\dot\phi^2\, ,
\end{equation}
\begin{equation}
\centering
\label{motion3}
\ddot\phi+3H\dot\phi+V'+\xi'\mathcal{G}=0\, ,
\end{equation}
where the condition $\ddot\xi=H\dot\xi$ has been used. As shown,
the equations themselves are quite simple compared to the
unconstrained model \cite{Pozdeeva:2020apf}. In addition, the fact
that the contribution of the Gauss-Bonnet term can be absorbed in
the prefactor of the Hubble rate expansion and its time derivative
in the Friedmann and Raychaudhuri equation respectively, in the
same manner simplifies the overall phenomenology, not just for
inflation but also for primordial eras in general. This is because
it can be treated as a general dynamical factor which does not
result in the inclusion of additional terms on the right hand
side, meaning the energy density, and can reach a fixed value for
de Sitter solutions, where no time dependence is expected. In
other words, the effective gravitational constant can be written
in the following form \cite{Linder:2021pek},
\begin{equation}
\centering
\label{Geff}
G_{eff}=\frac{G}{\alpha-64\pi G\ddot\xi}\, .
\end{equation}
This treatment of the factor $\alpha-8\kappa^2\ddot\xi$ has been
considered during the reheating era in Ref.
\cite{Venikoudis:2022gfg} where it was shown that the duration of
reheating is mildly affected by such term since it effectively
shifts the numerical value of the energy density $\rho_{end}$
without spoiling the condition $\dot\phi^2(t_{end})=V(\phi_{end})$
which applies to the case of a canonical scalar field. Now for
simplicity, in order to study the inflationary dynamics, since the
slow-roll assumption has already been made in the derivation of
$\dot\phi$ in Eq. (\ref{constraint2}), let us also use the
relations $\frac{1}{2}\dot\phi^2\ll V$ and
$8\kappa^2\ddot\xi\ll\alpha$. In turn, one finds that equations
(\ref{motion1}) and (\ref{motion2}) are written as,
\begin{equation}
\centering
\label{motion4}
H^2\simeq\frac{\kappa^2V}{3\alpha}\, ,
\end{equation}
\begin{equation}
\centering
\label{motion5}
\dot H\simeq-\frac{\kappa^2\dot\phi^2}{2\alpha}\, ,
\end{equation}
\begin{equation}
\centering \label{motion6} 3H\dot\phi+V'+24\xi'H^4\simeq0\, ,
\end{equation}
where the fact that $\dot H\ll H^2$ can easily be inferred from
the dynamics of the scalar field. Note also that Eq.
(\ref{motion3}) now, due to the fact that $\dot\phi$ is specified
from Eq. (\ref{constraint2}), can actually be treated as a
differential equation with respect to either the scalar potential,
which would then make it a first order differential equation, or
the Gauss-Bonnet scalar coupling function, thus making it a second
order differential equation. Regardless of the choice, only one
scalar function needs to be specified. By treating it as a second
order differential equation, the solution may be non-trivial as
integral forms can be extracted \cite{Venikoudis:2021irr}, however
integral forms can in fact be convenient as they suggest a
specific dependence of the tensor-to-scalar ratio or, as we shall
showcase subsequently the first slow-roll index, on the
$e$-foldings number $N$. This has been showcased explicitly in
Ref. \cite{Oikonomou:2022ksx} where depending on the relation
between $r$ and $N$, various integral forms were extracted by
means of reconstruction and by also using the constraint
$\ddot\xi=H\dot\xi$. Let us now proceed with the inflationary
phenomenology, and for the Einstein-Gauss-Bonnet model, the
spectral indices of the theory can be described properly if one
defines the following inflationary indices \cite{Hwang:2005hb},
\begin{align}
\centering
\label{slow-roll}
\epsilon_1&=-\frac{\dot H}{H^2},&\epsilon_2&=\frac{\ddot\phi}{H\dot\phi},&\epsilon_3&=\frac{\dot E}{2HE},&\epsilon_4&=\frac{\dot Q_t}{2HQ_t}\, ,
\end{align}
where
$E=\frac{\alpha}{\kappa^2}\bigg[1+\frac{3Q_a^2}{2Q_t\dot\phi^2}\bigg]$,
$Q_t=\frac{\alpha}{\kappa^2}+\frac{Q_a}{H}$, $Q_a=-8\dot\xi H^2$.
As shown, the third and fourth indices carry information about
string corrections, meaning they showcase the impact that the
Gauss-Bonnet density has on the overall phenomenology, while the
first indices are typical. Typically,
$Q_t=\frac{1}{\kappa_{eff}^2}$, therefore it is connected to the
shifted Planck mass due to the presence of the Gauss-Bonnet term.
While the slow-roll dynamics have been incorporated, the above
indices should not be considered as slow-roll indices because it
may be the case that their numerical value is actually large, see
Ref. \cite{Oikonomou:2022tux} where this is showcased explicitly
for an $f(R)$ model. Now by working with the previously extracted
expression of $\dot\phi$, one can show that the above indices take
the following approximate expressions,
\begin{equation}
\centering
\label{index1}
\epsilon_1=\frac{x}{1-2\beta x}\, ,
\end{equation}
\begin{equation}
\centering
\label{index2}
\epsilon_2=1-\epsilon_1-\gamma\, ,
\end{equation}
\begin{equation}
\centering
\label{index3}
\epsilon_3=\frac{3\beta^2\epsilon_1}{1+3\beta^2\epsilon_1}\bigg[1-2\epsilon_1-\epsilon_2-\epsilon_4\bigg]\, ,
\end{equation}
\begin{equation}
\centering
\label{index4}
\epsilon_4=-\beta\epsilon_1(1-\epsilon_1)\, ,
\end{equation}
where $x=\frac{1}{2\alpha}\bigg(\frac{\kappa\xi'}{\xi''}\bigg)^2$,
$\beta=8H^2\xi''$ and $\gamma=\frac{\xi'\xi'''}{\xi''^2}$.
Typically Eq. (\ref{constraint1}) suggests that $2\beta
x\to\frac{2\beta x}{1-\epsilon_2}$ in the denominator with
$x=\frac{1}{2\alpha}\bigg(\frac{\kappa\dot\phi}{H}\bigg)^2$, but
hereafter the latter is deemed subleading. Note that the condition
$8\kappa^2\ddot\xi\ll\alpha$ is not required in order to extract
the first index, but it was presented in order to specify the
Hubble rate (\ref{motion4}) and its derivative (\ref{motion5}). In
fact, by carrying out some calculations, it can be shown that
$\frac{8\kappa^2\ddot\xi}{\alpha}=\frac{2\beta\epsilon_1}{1+2\beta\epsilon_1}=2\beta
x$\footnote{In principle, given that $\epsilon_2$ participates in
Eq. (\ref{constraint2}), the result is $\frac{2\beta
x}{1-\epsilon_2}$ however $2\beta x\epsilon_2$ is deemed
subleading and is thus neglected from subsequent calculations}
therefore in the limit $\beta\epsilon_1\ll 1$, one obtains the
expression $\epsilon_1\simeq x$ which was used in Ref.
\cite{Odintsov:2020sqy} for $\alpha=1$. As shown, only $x$,
$\beta$ and $\gamma$ are needed in order to fully specify all the
indices, which in turn are specified by the free parameters of the
models and in particular $\xi(\phi)$ since $V(\phi)$ is connected
to $\xi(\phi)$ through Eq. (\ref{motion3}). Note that in order to
proceed, it is assumed that $\beta\epsilon_1\ll 1$. Let us now see
how the spectral indices are connected to the above results.
According to Ref. \cite{Hwang:2005hb}, the observables are now
functions of the indices in Eq. (\ref{slow-roll}) as follows,
\begin{align}
\centering
\label{spectralindices}
n_\mathcal{S}&=1-2\frac{2\epsilon_1+\epsilon_2+\epsilon_3}{1-\epsilon_1}\, ,&r&=16\bigg|\bigg(\epsilon_1-\frac{\kappa^2Q_e}{4\alpha H}\bigg)\frac{\alpha c_\mathcal{S}^3}{\kappa^2Q_t}\bigg|\, ,&n_\mathcal{T}&=-2\frac{\epsilon_1+\epsilon_4}{1-\epsilon_1}\, ,
\end{align}
where $Q_e=-32\dot\xi\dot H$ and $c_\mathcal{S}$ is the
propagation velocity of scalar perturbations. Here, it becomes
apparent that the tensor spectral index and the tensor-to-scalar
ratio are in fact specified only by $\epsilon_1$ and $\beta$ while
$\gamma$ can in principle affect the scalar spectral index. The
value of $\epsilon_2$ can become problematic in certain models if
it is quite large. Consider for instance the findings of Ref.
\cite{Oikonomou:2020oil} which showed that for a linear
Gauss-Bonnet scalar coupling function, $\epsilon_2=1$ identically
and therefore the scalar spectral index cannot become viable. This
is shown more transparently here since for
$\xi(\phi)=\frac{\phi}{f}$, Eq. (\ref{constraint2}) may not be
valid but the replacement
$-\frac{\beta\epsilon_1}{1+2\beta\epsilon_1}\to
y=-\frac{4\kappa^2\dot\phi H}{f\alpha}$ in equations
(\ref{index3}) and (\ref{index4}) while
$\epsilon_1=\frac{1}{2\alpha}\bigg(\frac{\kappa\dot\phi}{H}\bigg)^2$
and $\epsilon_2=1$ suffices. In consequence, the quite large value
of $\epsilon_2$ cannot be countered by anything else since
$\epsilon_1$, $\epsilon_3$ and $\epsilon_4$ remain small due to
the requirement that $\frac{\kappa\dot\phi}{H}\ll1$. It can be
shown that this feature can be avoided in extended modified
gravity theories, such as the $f(R)$ gravity case. Let us now
focus on the tensor spectral index and the tensor-to-scalar ratio.
By focusing on the constraint in Eq. (\ref{constraint1}), one can
show that,
\begin{align}
\centering
\label{tensorratio}
r&=16\epsilon_1c_\mathcal{S}^3\, ,
\end{align}
where it is simplified to quite an extent and in fact, for
$c_\mathcal{S}\simeq1$, it coincides with the usual result. Due to
the fact that $\epsilon_1$ is positive by definition, the absolute
value is lifted if we assume a positively defined sound wave
velocity which essentially reads \cite{Hwang:2005hb},
\begin{equation}
\centering
\label{cA}
c_\mathcal{S}^2=1-\frac{4\epsilon_1^2\beta^2}{1+3\beta^2\epsilon_1}\, ,
\end{equation}
and due to the fact that $\epsilon_1$ is considered to be small
because of the slow-roll assumption, it can easily be inferred
that by expanding $c_\mathcal{S}$ in terms of $\epsilon_1$, the
tensor-to-scalar ratio at leading order becomes,
\begin{equation}
\centering
\label{simplifiedr}
r=16\epsilon_1\bigg[1-6\beta^2\epsilon_1^2\bigg]+\mathcal{O}(\epsilon_1^4)\, .
\end{equation}
similar to the result extracted in Ref. \cite{Odintsov:2020zkl}
(in this case, it was assumed that $\frac{V'}{\kappa
V}=\tilde\gamma$ is a constant however it can easily be proven
that $\tilde\gamma\sim\sqrt{\epsilon_1}$ therefore the results
agree). Note that this expression is valid only if
$\beta\epsilon_1\ll 1$ otherwise Eq. (\ref{tensorratio}) should be
used. Once again, the case of a linear Gauss-Bonnet scalar
coupling function, although incompatible with observations, is
obtained by the replacement
$-\frac{\beta\epsilon_1}{1+2\beta\epsilon_1}\to y$. Overall, this
means that the tensor-to-scalar ratio is mildly shifted due to the
presence of the Gauss-Bonnet term, but it is mainly specified by
the first slow-roll index. Indeed, this has already been
considered in Ref. \cite{Oikonomou:2022ksx} where prior to the
reconstruction of $\xi(\phi)$, the tensor-to-scalar ratio was
considered to be linear in $\epsilon_1$, which as showcased here
is justifiable. Finally, it should be stated that the tensor
spectral index (\ref{spectralindices}) in the context of the
present formalism reads,
\begin{equation}
\centering
\label{simplifiednt}
n_\mathcal{T}=-\frac{2\epsilon_1}{1-\epsilon_1}+2\beta\epsilon_1\, ,
\end{equation}
if index $\epsilon_4$ is substituted from Eq. (\ref{index4}),
which can in fact become positive provided that
$\beta=8\xi''H^2>\frac{1}{1-\epsilon_1}$, or roughly speaking
$\beta>1$, which is in agreement with the findings of Ref.
\cite{Oikonomou:2021kql} where $1/\beta$ was essentially used. For
instance, having $\epsilon_1=0.0035$, $\beta=1.2$ and
$\gamma=1.001$, the observables are $n_\mathcal{S}=0.965038$,
$r=0.0559941$ and $n_\mathcal{T}=0.00137541$. Obviously the
Gauss-Bonnet term spoils the consistency relation as now
$r\neq-8n_\mathcal{T}$ unless $|\beta|\ll 1$. Note that for
consistency, since we require $8\ddot\xi\kappa^2\ll\alpha$, or in
other words $2\beta\epsilon_1\ll1$, the parameter $\beta$ in this
example was chosen to be of the order $\mathcal{O}(1)$ and
positive, but if it becomes negative, then there exists no
limitation to its value. It is also worth mentioning that the
above results clearly state that a blue-tilted tensor spectral
index cannot be derived for any model that predicts
$\xi''(\phi_{k})\leq 0$. The sign of the spectral index does not
affect the inflationary phenomenology, however a positive value is
connected to the amplification of the energy spectrum of
primordial gravitational waves, which in general is quantified by
the running Planck mass \cite{Oikonomou:2022xoq,Linder:2021pek},
\begin{equation}
\centering
\label{am}
a_M=\frac{2z}{1+2z}(1-\epsilon_1)\, ,
\end{equation}
with $z=-\frac{4\kappa^2\dot\xi H}{\alpha}$. This form is valid
for a random cosmological era however under the slow-roll
assumption one finds that
$z=-\frac{\beta\epsilon_1}{1+2\beta\epsilon_1}$ and thus
$a_M\simeq 2\epsilon_4$.

Lastly, it is worth mentioning in brief that for the case of
scalar non-Gaussianities, the equilateral non-linear term, similar
to the scalar spectral index, is affected by all the parameters
since, according to the findings of Ref. \cite{DeFelice:2011zh},
where
$f_{NL}^{eq}=\frac{55}{36}(\epsilon_1-4\delta_\xi)+\frac{5}{12}\eta+\frac{10}{3}\delta_\xi$
with $\eta=\frac{\dot\epsilon_1}{H\epsilon_1}$ and
$\delta_\xi=\frac{\kappa^2H\dot\xi}{\alpha}$, it can be shown that
now,
\begin{align}
\centering
\label{fNL}
f_{NL}^{eq}&=\frac{55}{36}\bigg[\epsilon_1-\beta x\bigg]+\frac{5}{6}\bigg[\epsilon_1+\epsilon_2+\beta x+\beta\epsilon_1(\gamma+2\epsilon_2)\bigg]\, ,
\end{align}
where in the limit $\xi\to 0$, the canonical scalar field result
is safely reached. For the linear Gauss-Bonnet scalar coupling
function, one would simply replace
$-\frac{\beta\epsilon_1}{1+2\beta\epsilon_1}$ with $y$ however, as
shown in \cite{Oikonomou:2020oil}, the model is not compatible
with the Planck data. Having a viable inflationary era with
$\epsilon_1,\epsilon_2\sim \mathcal{O}(10^{-3})$ such that the
scalar spectral index and the tensor-to-scalar ratio are
simultaneously in agreement with observations, suggests that the
equilateral non-linear term is approximately of the order
$\mathcal{O}(10^{-2}-10^{-3})$. This is expected since as shown in
Ref. \cite{DeFelice:2011zh}, only models with a small value for
the sound wave velocity $c_\mathcal{S}$ during the first horizon
crossing can result in relatively large values of the equilateral
non-linear term.  Therefore, for a potential driven inflation
model, the constrained Gauss-Bonnet model is essentially quite
similar to the canonical model case which seems to be in agreement
with the results of Ref. \cite{Odintsov:2020mkz} where the
constrained Gauss-Bonnet model was studied under the constant-roll
assumption.

\section{Running of the Spectral Indices in Rescaled Einstein-Gauss-Bonnet Models}

In the previous section it was showcased explicitly that the
spectral indices can we written as functions of the first
slow-roll index, and of  $\beta$ and $\gamma$, the last one
appearing only in the scalar spectral index. For a specific model,
after the free parameters of the model are specified, the final
value of the scalar field can be derived from the condition
$\epsilon_1(\phi_{end})=1$ which is indicative of the end of
inflation. Afterwards, the initial value of the scalar field can
be derived from the definition of the $e$-foldings number
$N=\int_{t_k}^{t_{end}}Hdt=\int_{\phi_k}^{\phi_{end}}\frac{H}{\dot\phi}d\phi$
which is the value that is inserted in the spectral indices and
the tensor-to-scalar ratio, in order for computing their numerical
values. In this approach, $\phi_k$ and $\phi_{end}$ are completely
specified by the free parameters of the model. The computation of
the aforementioned spectral indices is performed at the pivot
scale $k_*=0.05\,$Mpc$^{-1}$ and according to the latest Planck
data, the respective values are \cite{Planck:2018vyg},
\begin{align}
\centering
\label{spectralvalues}
n_\mathcal{S}&=0.9649\pm0.0042, &68\% \cl\, , \nonumber\\ r&<0.064, &95\% \cl\, ,
\end{align}
while the tensor spectral index remains unspecified given that
$B$-modes have yet to be observed in the CMB
\cite{Kamionkowski:2015yta}, however an upper bound exists,
provided that a blue-tilted tensor spectral index is obtained,
which suggests that $n_\mathcal{T}<0.5$, see Ref.
\cite{Giare:2020vss}, since values beyond this threshold cannot be
the result of a stochastic gravitational wave background. These
values are nearly scale invariant, meaning that changing the
wavenumber $k$ should not in principle alter the numerical value
of the spectral indices computed at the pivot scale $k_*$. By
definition, the tensor-to-scalar ratio is specified at the pivot
scale, so there exists no reason to evaluate it at different
scales, however the spectral indices, as functions of $k$, can be
computed for a plethora of values for $k$ differing from the CMB
pivot scale. In fact, by expanding the power-spectra, one can show
that the spectral indices scale indeed with $k$ as shown below
\cite{Zarei:2014bta},
\begin{equation}
\centering
\label{running}
n_{\mathcal{S}/\mathcal{T}}(k)=n_{\mathcal{S}/\mathcal{T}}(k_*)+\sum_{n=1}^\infty\frac{d^nn_{\mathcal{S}/\mathcal{T}}}{d\ln k^n}\bigg|_{k_*}\frac{\ln^n\frac{k}{k_*}}{(n+1)!}\, ,
\end{equation}
which is a power-series. As shown, higher powers of $n$ do not
become so important and thus one can focus solely on the first
derivatives
$a_{\mathcal{S}/\mathcal{T}}(k_*)=\frac{dn_{\mathcal{S}/\mathcal{T}}}{d\ln
k}\bigg|_{k_*}$. Recent observations speculate that their
numerical values vary between $\mathcal{O}(10^{-4}-10^{-3})$
\cite{Planck:2018vyg}, therefore it would be interesting to see
how the constraint in Eq. (\ref{constraint1}) can in fact affect
the running of the spectral indices. We shall perform the
computations analytically and derive conclusions based on the
example mentioned previously that manifests a blue-tilted tensor
spectral index.

Let us start by computing the corresponding expression of
$a_\mathcal{T}(k_*)$. According to Eq. (\ref{simplifiednt}) the
running of the tensor spectral index for an arbitrary
Einstein-Gauss-Bonnet model at first order, using the above
auxiliary dimensionless parameters, should be,
\begin{widetext}
\begin{equation}
\centering
\label{ntrun}
a_\mathcal{T}(k_*)=\frac{1}{1-\epsilon_1}\bigg[\bigg(2(\epsilon_1+\epsilon_2)+2\beta\epsilon_1\bigg(\gamma+2\epsilon_2\bigg)\bigg)\bigg\{n_\mathcal{T}-2\bigg(\frac{\epsilon_1}{1-\epsilon_1}\bigg)^2\bigg\}+2\beta\epsilon_1(\gamma-2\epsilon_1)\bigg]\, ,
\end{equation}
\end{widetext}
so roughly speaking it is of the order $\mathcal{O}(10^{-3})$ and
furthermore, in agreement with estimates \cite{Planck:2018vyg}.
Here, we made use of the chain-rule $\frac{dn_\mathcal{T}}{d\ln
k}=\frac{dn_\mathcal{T}}{dN}\frac{dN}{d\ln k}$ with the respective
derivatives being $\frac{dN}{d\ln k}=\frac{1}{1-\epsilon_1}$, as
is the case with most $f(R,\phi)$ models, and
$\frac{d}{dN}=\frac{1}{H}\frac{d}{dt}$ is the differential
operator constructed according to the definition of the
$e$-foldings number. Note that in the limit $\beta x\ll 1$,
$\epsilon_1\simeq x$ therefore several simplifications occur. In
addition, for the interesting case of $\beta>1$, it is possible to
obtain a positive value.

Overall, the expressions that were used are,
\begin{align}
\centering
\label{derivatives}
\frac{\dot\epsilon_1}{H\epsilon_1}&=2(\epsilon_1+\epsilon_2)+2\beta \epsilon_1(\gamma+2\epsilon_2)\, ,&\frac{\dot\beta}{H\beta}&=\gamma-2\epsilon_1\, ,
\end{align}
with the first holding true for most scalar models while the
second appears in the Einstein-Gauss-Bonnet case. Let us now
proceed with the derivation of $a_\mathcal{S}(k_*)$. In general,
for the scalar spectral index, one requires additional
information. If the variation of $\gamma$, which participates in
(\ref{index2}), is extracted, we find that,
\begin{equation}
\centering
\label{gamma}
\frac{\dot\gamma}{H\gamma}=1-2\gamma+\delta\, ,
\end{equation}
with $\delta$ being a new parameter
$\delta=\frac{\xi'\xi''''}{\xi''\xi'''}$. This is a natural
consequence as further derivatives of the Gauss-Bonnet scalar
coupling function should appear in the perturbed equations,
similar to the case of the tensor spectral index mentioned before.
The only condition is that the fourth derivative of $\xi(\phi)$
exists, otherwise this term can be discarded, for instance in a
$\xi\sim \phi^3$ model, however a well behaved $\gamma$, such that
a compatible with observations $n_\mathcal{S}$ is extracted,
suggests that $\delta\simeq \gamma$.  In consequence, one can show
that the evolution of the second index $\epsilon_2$ with respect
to the $e$-foldings number is,
\begin{equation}
\centering
\label{e2ratio}
\frac{\dot\epsilon_2}{H}=-2\epsilon_1(\epsilon_1+\epsilon_2)-2\beta \epsilon_1^2(\gamma+2\epsilon_2)-\gamma(1-2\gamma+\delta)\, .
\end{equation}
Similarly, it can easily be inferred that the fourth index
(\ref{index4}) evolves as,
\begin{align}
\centering
\label{e4}
\frac{\dot\epsilon_4}{H}&=\epsilon_4(\gamma+2\epsilon_2)(1+2\beta\epsilon_1)+2\beta\epsilon_1^2\bigg[\epsilon_1+\epsilon_2+\beta\epsilon_1(\gamma+2\epsilon_2)\bigg]\, ,
\end{align}
therefore, the evolution of the third (\ref{index3}) and most
complex index can be specified by the following and quite lengthy
expression,
\begin{align}
\centering
\label{e3ratio}
&\frac{\dot\epsilon_3}{H}=\frac{1}{1+3\beta^2\epsilon_1}\bigg[\epsilon_3\bigg(2\frac{\dot\beta}{H\beta}+\frac{\dot\epsilon_1}{H\epsilon_1}\bigg)-3\beta^2\epsilon_1\bigg(2\frac{\dot\epsilon_1}{H}+\frac{\dot\epsilon_2}{H}+\frac{\dot\epsilon_4}{H}\bigg)\bigg]\, .
\end{align}
These results are extracted without invoking any simplifications
apart from the condition (\ref{constraint2}), in which the second
index $\epsilon_2$ is absent since it was assumed that
$\epsilon_2\ll 1$, for simplicity however typically the second
term in both expressions which scales as
$\mathcal{O}(\epsilon_1^3)$ is subleading. In the end, if we
combine all the above expressions, then the running of the scalar
spectral index reads,

\begin{align}
\centering \label{dns}
a_{\mathcal{S}}(k_*)&=\frac{-2}{(1-\epsilon_1)^2}\bigg\{2\epsilon_1\bigg(1+\frac{1-n_{\mathcal{S}}}{2}\bigg)\bigg[\epsilon_1+\epsilon_2+\beta
\epsilon_1(\gamma+2\epsilon_2)\bigg]-\gamma(1-2\gamma+\delta)+\frac{1}{1+3\beta^2\epsilon_1}\bigg[2\epsilon_3\bigg(\epsilon_2-\epsilon_1+\gamma+\beta\epsilon_1(\gamma+2\epsilon_2)\bigg)\nonumber\\&-3\beta^2\epsilon_1\bigg(2\epsilon_1(\epsilon_1+\epsilon_2+\beta
\epsilon_1(\gamma+2\epsilon_2))-\gamma(1-2\gamma+\delta)+\epsilon_4(1+2\beta\epsilon_1)(\gamma+2\epsilon_2)+2\beta\epsilon_1^2(\epsilon_1+\epsilon_2+\beta\epsilon_1(\gamma+2\epsilon_2))\bigg)\bigg]\bigg\}\,
,
\end{align}

from which the leading order seems to be around
$\mathcal{O}(10^{-3})$ provided that $\delta$ is well behaved.
Further simplifications occur in the limit $\beta x\ll 1$ where
$\epsilon_1\simeq x$. Furthermore, letting $\xi\to 0$ suggests
that
$a_{\mathcal{S}}(k_*)\simeq-4\epsilon_1(\epsilon_1+\epsilon_2)$
while $a_{\mathcal{T}}(k_*)\simeq
2(\epsilon_1+\epsilon_2)n_\mathcal{T}$ which are the expected
forms for the canonical scalar field case \footnote{For the scalar spectral index, the prefactor is 4 instead of 12 since $\epsilon_2$ in this approach scales with $-\epsilon_1$.} and are both negative.
Overall, for viable inflationary models that seem to manifest
spectral indices in agreement with observations, the running of
the spectral indices is also negligible thus justifying the
statement that the spectral indices are nearly scale invariant.
The fact that a blue-tilted tensor spectral index is a plausible
scenario for a relatively large value for parameter $\beta$ or
that the model is constrained to satisfy the relation
$\ddot\xi=H\dot\xi$ throughout the evolution of the universe does
not seem to spoil the expected results \cite{Planck:2018vyg} for
$a_{\mathcal{S}/\mathcal{T}}(k_*)$.

\section{Swampland Criteria for Einstein-Gauss-Bonnet Models in agreement with the GW170817 Event}

In this section of this paper we shall briefly discuss the
validity of the Swampland criteria and the circumstances under
which a viable inflationary model may reside in the Swampland. The
Swampland criteria
\cite{Vafa:2005ui,Ooguri:2006in,Palti:2020qlc,Brandenberger:2020oav,Blumenhagen:2019vgj,Wang:2019eym,Benetti:2019smr,Palti:2019pca,Cai:2018ebs,Akrami:2018ylq,Mizuno:2019pcm,Aragam:2019khr,Brahma:2019mdd,Mukhopadhyay:2019cai,Brahma:2019kch,Haque:2019prw,Heckman:2019dsj,Acharya:2018deu,Elizalde:2018dvw,Cheong:2018udx,Heckman:2018mxl,Kinney:2018nny,Garg:2018reu,Lin:2018rnx,Park:2018fuj,Olguin-Tejo:2018pfq,Fukuda:2018haz,Wang:2018kly,Ooguri:2018wrx,Matsui:2018xwa,Obied:2018sgi,Agrawal:2018own,Murayama:2018lie,Marsh:2018kub,Storm:2020gtv,Trivedi:2020wxf,Sharma:2020wba,Odintsov:2020zkl,Mohammadi:2020twg,Trivedi:2020xlh,Han:2018yrk,Achucarro:2018vey,Akrami:2020zfz},
see also \cite{Colgain:2018wgk,Colgain:2019joh,Banerjee:2020xcn},
specify whether a model is in fact UV incomplete or not, depending
on whether they are satisfied or not, and suggest that,
\begin{itemize}
\item$|\kappa\Delta\phi|\leq\mathcal{O}(1)$. This is the Swampland Distance Conjecture. It suggests that as an effective theory, a specific field range exists therefore the difference between the initial and final value of the scalar field during inflation cannot be arbitrarily large, it is however independent of the sign.
\item$\bigg|\frac{V'}{\kappa V}\bigg|\geq\mathcal{O}(1)$. The second condition is the de Sitter conjecture. It suggests that for a positively defined scalar potential, its derivative with respect to the scalar field at the start of inflation must have a lower bound but its sign is once again irrelevant.
\end{itemize}
It should be stated that there exists also a third criterion,
mostly used as a supplementary condition for the second criterion,
which is connected to the second derivative of the scalar
potential as $-\frac{V''}{\kappa^2V}\geq\mathcal{O}(1)$ and states
that the second derivative of the scalar potential during the
first horizon crossing is not only negative but it also has a
lower bound. The aforementioned criteria that distinguish models
depending on their UV completeness are not mandatory and do not
need to be satisfied simultaneously in order for a model to be UV
incomplete. The easiest example that one can consider is the
power-law model, see for instance Ref. \cite{Oikonomou:2021zfl}
where it was shown that the chaotic model cannot satisfy
simultaneously the second and third criterion since they are
contradictory. In general, even if one condition is met then the
criteria are satisfied and thus the theory can be considered as UV
incomplete.

For the sake of simplicity, in the current paper we shall focus
our work mainly on the second criterion and its phenomenological
implications, however the rest shall be briefly considered. Let us
commence by combining equations (\ref{constraint2}) and
(\ref{motion3}) which, upon performing a few calculations, it can
easily be inferred that at the moment where modes become
superhorizon, the second criterion can be written approximately
with respect to the previously defined auxiliary parameters as,
\begin{equation}
\centering
\label{swamp1}
\frac{V'}{\kappa V}=-\frac{3H^2}{\kappa^2V}\frac{\kappa\xi'}{\xi''}\bigg[1+\frac{\epsilon_2}{3}+\beta(1-\epsilon_1)\bigg]\simeq-\frac{1}{\alpha}\frac{\kappa\xi'}{\xi''}\bigg[1+\beta\bigg]\, ,
\end{equation}
from which it becomes clear that, due to the fact that the
criterion is proportional to the square-root of the first
slow-roll index (\ref{index1}), given that
$\sqrt{\alpha\epsilon_1}\sim|\frac{\kappa\xi'}{\xi''}|$, it
requires either a large value for $\beta$, regardless of its sign,
or a quite small value for $\alpha$, if not both. Indeed, it can
be shown that for a small value of the parameter $\alpha$, in
particular for $\alpha\sim\mathcal{O}(10^{-3})$, the second
condition is met since its numerical value increases beyond
$\mathcal{O}(1)$. Note also that $\epsilon_2$ is not included,
however in principle it participates in Eq. (\ref{constraint2}),
therefore it should manifest here as well, it proves however to be
subleading. Conclusions can be drawn for the third criterion as
well since a derivative of (\ref{motion3}) for a subleading
$\ddot\phi$ with respect to the scalar field suggests that,
\begin{align}
\centering
\label{swamp2}
-\frac{V''}{\kappa^2V}&=\frac{3H^2}{\kappa^2V}\bigg[(\epsilon_2-\epsilon_1)(1+\frac{\epsilon_2}{3})+\beta(1-\epsilon_1)(1-4\epsilon_1)+\frac{\dot\epsilon_2}{3H}-\beta\frac{\dot\epsilon_1}{H}\bigg]\simeq\frac{1}{\alpha}\bigg[\epsilon_2-\epsilon_1+\beta(1-5\epsilon_1)\bigg]\, ,
\end{align}
which is in agreement with the previous statements. In both cases,
Eq. (\ref{motion4}) was used however if
$1-\frac{8\kappa^2\ddot\xi}{\alpha}$ is considered to participate
in the Friedmann equation then a factor of $1+2\beta\epsilon_1$
should appear in equations (\ref{swamp1}) and (\ref{swamp2}) with
$\frac{1}{2}\dot\phi^2$ being neglected. In this case, certain
variations of $\beta$, $x$ and $\epsilon_1$ should appear in Eq.
(\ref{swamp2}) and if one were to include the second slow-roll
index in Eq. (\ref{constraint2}), more perplexed expressions would
emerge but they should be regarded as subleading corrections.
Therefore, it is possible to satisfy the Swampland criteria in the
constrained Einstein-Gauss-Bonnet model, even though the second
condition scales as $\sqrt{\epsilon_1}$, according to the
continuity equation of the scalar field and the third is linear in
$\epsilon_1$. Note that for $\beta>1$ and $\alpha=1$, the third
criterion is satisfied along with $n_\mathcal{T}>0$ therefore it
is possible to satisfy the Swampland criteria while a blue
spectrum is obtained. Furthermore, the Lyth bound
\cite{Lyth:1996im} which reads $\sqrt{\frac{\alpha r}{8}}\Delta
N<|\kappa\Delta\phi|<\mathcal{O}(1)$ can be satisfied provided
that the ratio $\frac{\kappa\xi'}{\xi''}$ is in fact quite small
during the first horizon crossing, something which is guaranteed
if the inflationary phenomenology is actually viable. As a result,
we find that the Lyth bound at first order, according to Eq.
(\ref{simplifiedr}) reads \cite{Odintsov:2020zkl}
$|\kappa\Delta\phi|>\Delta N|\frac{\kappa\xi'}{\xi''}|=\Delta
N\sqrt{2\alpha x}$ and for $\Delta N\sim60$, one can estimate that
$|\frac{\kappa\xi'}{\xi''}|$ must be of order
$\mathcal{O}(10^{-3})$ and lesser. Note that the result is
actually independent of the parameter $\alpha$. Typically at first
order, one should require $\epsilon_1\leq 0.004$ in order for Eq.
(\ref{simplifiedr}) to be in agreement with
(\ref{spectralvalues}). While the compatibility of the scalar
spectral index (\ref{spectralindices}) may seem to be spoiled due
to the fact that $\epsilon_1$ is small,  the linear combination of
indices (\ref{index2}) and (\ref{index3}) may still result in
viable results for $n_\mathcal{S}$ while respecting the upper
bound for $r$. Hence, having a small value for $\alpha$ and
satisfying the condition $\frac{\kappa\xi'}{\xi''}\ll1$ suffices
in order to satisfy most, if not all, Swampland criteria.
Therefore, in a sense, the Lyth-bound that imposes an upper bound
on the generation of primordial gravitational waves now manages to
control the ratio between the first two derivatives of the
Gauss-Bonnet scalar coupling function during the first horizon
crossing for a potential driven inflationary model. Hence, it
becomes clear that for a small value of $\alpha$ where the
Swampland criteria are satisfied, the Lyth-bound can also be in
agreement with the Swampland since the bound is controlled by the
fraction $\frac{\kappa\xi'}{\xi''}$ and is strongly affected by
$\alpha$.

As a final note, let us briefly consider solutions that respect
the tracking condition \cite{Steinhardt:1999nw}. They were
initially introduced in quintessence theories and essentially
describe universal inflationary attractors to which the solution
converges for a wide range of initial conditions. In order for
this to occur, the following relation must be respected,
\begin{equation}
\centering
\label{tracking1}
 \frac{V'}{V}=\frac{H}{\dot\phi}\, ,
\end{equation}
which relates the ratio between the scalar potential and its
derivative with the derivative of the $e$-foldings number. This
can easily be inferred from the fact that
$N=\int_{t_k}^{t_{end}}Hdt=\int_{\phi_k}^{\phi_{end}}\frac{H}{\dot\phi}d\phi$
for a scalar model. Usually, the tracking condition cannot be
satisfied because it is at variance with the slow-roll dynamics.
This can be seen from that fact that while $V'\sim H\dot\phi$ in
usual slow-roll inflation models with a canonical scalar field,
the denominator implies that $V\sim\dot\phi^2$ however the
slow-roll conditions forbid this equivalence. As a result, the
tracking condition cannot be satisfied simultaneously with the
slow-roll conditions and one needs to resort to different ways
that may make Eq. (\ref{tracking1}) a viable relation, see for
instance Ref. \cite{Oikonomou:2021yks} where the constant-roll
assumption is invoked. For the case of a Gauss-Bonnet model
however, the ratio $\frac{H}{\dot\phi}$ is well known and is in
fact connected to the Gauss-Bonnet scalar coupling function
through the relation (\ref{constraint2}). By performing this
substitution and working on the general expression, one can show
that,
\begin{equation}
\centering
\label{tracking2}
\frac{V'}{\kappa V}=\frac{\xi''}{\kappa\xi'}\, ,
\end{equation}
which needs to be large due to the fact that
$\frac{\kappa\xi'}{\xi''}\ll1$, which is also in agreement with
the Lyth bound. In other words, when working on constrained
Einstein-Gauss-Bonnet models which are in agreement with the
GW170817 event, the tracking condition may be satisfied only if
the Swampland criteria are simultaneously satisfied, similar to
the results of the constant-roll case which were presented in Ref.
\cite{Oikonomou:2021yks}. Keep in mind that while it is possible
for the Swampland criteria to be satisfied without the tracking
condition, the opposite is not an option. Also, this result
applies to the case of additional string corrections that do not
affect the propagation velocity of tensor perturbations in Eq.
(\ref{cT}), for instance the Galilean model
$\xi(\phi)\Box\phi\nabla_\mu\phi\nabla^\mu\phi$ or additional
kinetic terms $\xi(\phi)(\nabla_\mu\phi\nabla^\mu\phi)^2$. This is
an interesting observation since it relates the UV completeness of
the model with the existence of universal inflationary attractors.
In addition, by taking the derivative of (\ref{tracking1}) with
respect to $\phi$ and neglecting $\epsilon_2$, one finds that,
\begin{equation}
\centering
\label{tracking3}
\frac{V''}{\kappa^2V}=\bigg(\frac{V'}{\kappa V}\bigg)^2\bigg[1-\epsilon_1-\epsilon_2\bigg]\, ,
\end{equation}
from which it becomes abundantly clear that while its numerical
value is large, its sign is positive, therefore the third
condition is not satisfied. This in principle is not a problem
since as it was stated before, the criteria are met even if only
one condition is satisfied. Therefore, the tracking condition
simply states that the third criterion is not satisfied however if
the sign is not important but only the order of magnitude is, then
both are satisfied alongside the tracking condition. This is not
in contrast to the previously extracted results as it was
showcased before that $\frac{V'}{\kappa V}\sim\sqrt{\epsilon_1}$
while $\frac{V''}{\kappa^2V}\sim\epsilon_1$. Furthermore, by
combining equations (\ref{swamp1}) and (\ref{tracking2}), one can
show that parameter $\beta$ should be fixed as follows,
\begin{equation}
\centering
\label{betavalue}
\beta\simeq-\frac{1}{2\epsilon_1}\, ,
\end{equation}
which works only if $\xi''(\phi_k)<0$ since it was assumed that
$\alpha$ is positive. The same result is also extracted if one
replaces (\ref{motion3}) in (\ref{tracking1}). The fact that
$\beta<0$ implies that there exists no Gauss-Bonnet model that
manages to respect the tracking condition while it simultaneously
produces a blue-tilted tensor spectral index.  In fact, if one
substitutes this value of $\beta$ in the tensor spectral index
(\ref{spectralindices}), it becomes apparent that the value
is\footnote{Keeping the next leading order terms in the expression
of $\beta$ alters the prefactor of $\epsilon_1$ from 2 to 5.}
$n_\mathcal{T}\simeq-1-2\epsilon_1$ provided that
$\epsilon_{1}\leq\mathcal{O}(10^{-3})$. Hence, satisfaction of the
Swampland criteria cannot be connected to a fixed sign for the
tensor spectral index. The result is the same regardless of
whether Eq. (\ref{motion1}) or (\ref{motion4}) is used in
(\ref{swamp1}). As stated before, the fact that a red-tilted
tensor spectral index is manifested simply suggests that the
energy spectrum of primordial gravitational waves cannot be
amplified \cite{Oikonomou:2022xoq}, the inflationary phenomenology
however remains viable. Note also that by equating (\ref{swamp2})
with (\ref{tracking3}) and using (\ref{betavalue}), a constraint
on $\epsilon_2$, or in other words $\gamma$, is imposed however in
this case there seems to be no agreement with the scalar spectral
index. Due to the large value of $\beta$, $\gamma$ needs to be
also positive and larger than unity however this assumption breaks
the approximation imposed in Eq. (\ref{constraint2}) so a
different approach in the constrained model is needed.

Before we closing this section, let us briefly comment on a
possible expression that the scalar functions may share. Suppose
that the Gauss-Bonnet scalar coupling function is connected to the
scalar potential through the relation,
\begin{equation}
\centering
\label{xiswamp}
\xi(\phi)=\frac{\lambda}{\kappa^4V(\phi)}\, ,
\end{equation}
where $\lambda$ is an auxiliary dimensionless parameter introduced
for the sake of generality. This form is usually proposed since it
serves as a quintessence model with viable late-time, therefore it
would be interesting to see what the tracking condition implies
for this designation. The above constraint does not spoil the
criteria since $\frac{\xi''}{\kappa\xi'}$ is large and this is
given by the second and third condition. By taking consecutive
derivatives with respect to the scalar field and substituting in
the tracking condition, the following expression is derived,
\begin{equation}
\centering
\label{swamprelation}
3\bigg(\frac{V'}{\kappa V}\bigg)^2=\frac{V''}{\kappa^2V}\, ,
\end{equation}
where as shown, since the second criterion is satisfied, the third
has a quite large value, however the opposite sign and is thus not
respected. This is the result of a power-law model as in the case
of Ref. \cite{Odintsov:2020sqy} and it can be proved analytically
if Eq. (\ref{swamprelation}) is treated as a differential equation
and not consider it as being valid only during the first horizon
crossing. Upon solving this differential equation, the scalar
potential reads, $V(\phi)=\frac{V_0}{\sqrt{\kappa(\phi+\varphi)}}$
where $V_0$ is the potential amplitude and $\varphi$ is the
integration constant with mass dimensions of eV for consistency.
This in turn fixes the Gauss-Bonnet scalar coupling function into
a power-law form with inverse power, that is however smaller than
unity as now $\xi(\phi)=\tilde\lambda\sqrt{\kappa(\phi+\varphi)}$
with $\tilde\lambda=\frac{\lambda}{\kappa^4V_0}$. If the initial
value of the scalar field is extracted exactly as it was mentioned
at the start of the previous section then we find that
$\phi_k\sim\e^{2N}M_{Pl}$ which obviously explodes for large values
of the $e$-foldings number, and thus it cannot produce a viable
inflationary phenomenology. Another way that may convince the
reader it the comparison of (\ref{tracking3}) with
(\ref{swamprelation}) which, although they should approximately
agree to some extend, it is impossible since the slow-roll indices
$\epsilon_1$ and $\epsilon_2$ have quite small values. Therefore,
the condition $\xi(\phi)V(\phi)=\lambda M_{Pl}^4$ cannot result in a
viable inflationary era even though it respects the tracking
condition. The same applies to the linear connection
$\xi(\phi)=\lambda\kappa^4V(\phi)$ which, upon substituting in the
tracking conditions, it generates the following relation,
\begin{equation}
\centering
\label{swamprelation2}
\bigg(\frac{V'}{\kappa V}\bigg)^2=\frac{V''}{\kappa^2V}\, ,
\end{equation}
which once again implies that the third condition is at variance
with the Swampland criteria, while the second is in agreement,
however the solution of this differential equation is an
exponential function $V(\phi)=V_0\e^{-\frac{\phi}{f}}$ and this
coupling choice cannot describe the inflationary era properly as
the first slow-roll index becomes $\phi$ independent, therefore
one obtains a description for eternal or no inflation at all
depending on the magnitude of the exponent
\cite{Odintsov:2020sqy}. Note also that this condition is
identical to Eq. (\ref{tracking3}).

Before closing this section, let us briefly address a somewhat
useful issue related to the above calculations. In principle, the
proper solution of the constraint $\ddot\xi=H\dot\xi$ suggests
that the second slow-roll index is,
\begin{equation}
\centering \label{eq1} \epsilon_2=1-\frac{\xi''}{H\xi'}\dot\phi\,
,
\end{equation}
which in turn implies that continuity equation of the scalar field
produces the following solution,
\begin{equation}
\centering \label{eq2}
\dot\phi=\frac{2H\xi'}{\xi''}\bigg[1\pm\sqrt{1+\frac{\xi''}{4H^2\xi'}(V'+\xi'\mathcal{G})}\bigg]\,
.
\end{equation}
 In other words, one could argue that,
\begin{equation}
\centering \label{eq3}
\epsilon_2=-1+2\sqrt{1-\frac{3}{4}\frac{\dot\phi_{UC}}{\dot\phi_C}}\,
,
\end{equation}
where $UC$ and $C$ subscripts refer to the unconstrained and
constrained expressions for $\dot\phi$ extracted under the
slow-roll assumption, meaning that
$\dot\phi_{UC}=-\frac{V'+\xi'\mathcal{G}}{3H}$ and
$\dot\phi_C=\frac{H\xi'}{\xi''}$. Hence, the continuity equation,
without any approximations, suggests that,
\begin{equation}
\centering \label{eq4} \frac{V'}{\kappa
V}=-\frac{1}{\alpha}\frac{\kappa\xi'}{\xi''}\bigg[\frac{1+\frac{\dot\phi^2}{2V}}{1-\frac{8\kappa^2\dot\xi
H}{\alpha}}\bigg]\bigg[\frac{\dot\phi_{UC}}{\dot\phi_C}+\beta(1-\epsilon_1)\bigg]\,
,
\end{equation}
while the tracking condition suggests that $\frac{V'}{\kappa
V}=\frac{H}{\kappa\dot\phi}$. In the end, by equating the two
expressions, the solution for $\beta$ is,
\begin{equation}
\centering \label{eq5}
\beta=-\frac{1}{1-\epsilon_1}\bigg[\frac{\dot\phi_{UC}}{\dot\phi_C}+\frac{1-\frac{8\kappa^2\dot\xi
H}{\alpha}}{1+\frac{\dot\phi^2}{2V}}\frac{\alpha}{2}\bigg(\frac{\xi''}{\kappa\xi'}\bigg)^2\frac{1}{1-\sqrt{1-\frac{3}{4}\frac{\dot\phi_{UC}}{\dot\phi_C}}}\bigg]\,
,
\end{equation}
which remains negative, therefore indeed the spectrum cannot be
blue even if the complete expression is considered. Of course, the
fact that the slow-roll condition is violated, simply states that
the scalar spectral index cannot obtain a phenomenologically
acceptable value, therefore the tracking condition is not a viable
option. There exists a difference since $\dot\phi$ and
$\dot\phi_C$ differ by a factor of 2 but is approximately the same
if $\frac{3}{4}\frac{\dot\phi_{UC}}{\dot\phi_C}$ is small. In
principle, $\beta$ in this approach can become positive for large
$\frac{\dot\phi_{UC}}{\dot\phi_C}$ but then the second slow-roll
index becomes also large.

\section{Preheating Era in Rescaled Einstein-Gauss-Bonnet models}

In this section we shall briefly discuss the phenomenological
implications that the constraint in Eq. (\ref{constraint1}) has on
the preheating era. This will be done by following the same steps
as in Ref. \cite{Venikoudis:2022gfg}. Typically, preheating occurs
immediately after inflation ceases in order to prepare the
conditions so that the Universe can be thermalized. This is
necessary, given that the temperature decreases drastically during
inflation owning to the quasi-de Sitter expansion. In order to
extract information, we relate the frequency of a mode at the
pivot scale $k_*$ with the current value of the Hubble rate
expansion by considering intermediate cosmological eras of
interest.  In particular \cite{Cook:2015vqa},
\begin{equation}
\centering \label{kmode}
\frac{k_*}{a_0H_0}=\frac{a_k}{a_{end}}\frac{a_{end}}{a_{pre}}\frac{a_{pre}}{a_{re}}\frac{a_{re}}{a_{eq}}\frac{a_{eq}}{a_0}\frac{H_{eq}}{H_0}\frac{H_k}{H_{eq}}\frac{1}{c_{\mathcal{S}}}\,
,
\end{equation}
where subscripts ``pre'' and ``re'' refer to the end of preheating
and reheating era respectively, while ``eq'' denotes the
matter-radiation equivalence. Typically, during the first horizon
crossing, the relation $c_{\mathcal{S}}k_*=a_kH_k$ holds true
hence the reason why it appears in the denominator however,
provided that $c_{\mathcal{S}}\simeq1$ at the start of inflation,
it is usually omitted in the literature. Here, for the sake of
consistency, we shall keep its contribution in the final
expression for the duration of the preheating era. Now by
recalling that the $e$-foldings number is defined as $N=\ln a$, it
can easily be inferred that,
\begin{equation}
\centering \label{logk}
\ln\frac{k_*}{a_0H_0}=-N_k-N_{pre}-N_{re}+\ln\frac{a_{re}}{a_0}+\ln\frac{H_k}{H_0}-\ln
c_{\mathcal{S}}\, ,
\end{equation}
where $N_k=\ln\frac{a_{end}}{a_k}$,
$N_{pre}=\ln\frac{a_{pre}}{a_{end}}$ and
$N_{re}=\ln\frac{a_{re}}{a_{pre}}$ are the duration of inflation,
preheating and reheating respectively. Now in order to proceed, we
shall exploit the fact that the Universe is evolving
adiabatically, therefore the scale factor at the end of reheating
and the current value are connected to the inverse of the
temperature, or in other words $T\sim\frac{1}{a}$. In the end, one
can show that \cite{Cook:2015vqa},
\begin{equation}
\centering \label{entropyconservation}
\frac{a_{re}}{a_0}=\frac{T_0}{T_{re}}\bigg(\frac{43}{11g_*}\bigg)^{\frac{1}{3}}\,
,
\end{equation}
where $g_*$ denotes the relativistic degrees of freedom in the
radiation domination era. Until now, the reheating temperature is
not known, however due to the fact that at the start of the
radiation domination era the Universe has reached thermal
equilibrium, it can easily be inferred that,
\begin{equation}
\centering \label{redensity}
\frac{3H^2_{re}}{\kappa^2}=\rho_{re}=\frac{\pi^2g_*}{30}T_{re}^4\,
.
\end{equation}
This expression can be used in order to replace $T_{re}$ in Eq.
(\ref{entropyconservation}) however $\rho_{re}$ is still unknown.
In order to specify it, we shall assume that the EoS between the
preheating and reheating era is approximately the same. Suppose
also that $\zeta$ is an auxiliary dimensionless parameter that
relates the energy density at the start of preheating with that at
the end of preheating, that is \cite{ElBourakadi:2021nyb},
\begin{equation}
\centering \label{zeta} \rho_{end}=\zeta \rho_{pre}\, ,
\end{equation}
where it is understood that approximately $\rho_{end}\sim
a_{end}^{-3(1+\omega_{pre})}$ with $\omega_{pre}$ residing in the
area $(-\frac{1}{3},1]$. Note that the effective EoS is assumed to
be arbitrary so that a non-canonical reheating era can be
realized, however it obtains values inside this interval in order
to obtain a decelerating expansion. This can easily be understood
since the special value of $\omega_{pre}=-\frac{1}{3}$ corresponds
to $\ddot a=0$ while $\omega_{pre}=1$ is the maximally allowed
value that respects causality. In consequence, the energy density
at the start of the radiation dominance can be extracted from the
end of the inflationary era as,
\begin{equation}
\centering \label{rhore}
\rho_{end}=\zeta\rho_{re}\bigg(\frac{a_{pre}}{a_{re}}\bigg)^{-3(1+\omega_{pre})}\,
,
\end{equation}
where for consistency, $\zeta\gg 1$. Now typically, one can
extract information about the energy density at the end of
inflation by postulating that $\dot H=-H^2$. By using equations
(\ref{motion1}) and (\ref{motion2}), it becomes clear that the
condition extracted is,
\begin{equation}
\centering \label{equivalence} \dot\phi^2_{end}=V_{end}\, ,
\end{equation}
which is exactly the same as in the canonical scalar field case.
As a result, the energy density reads \cite{Venikoudis:2022gfg},
\begin{equation}
\centering \label{rhoend}
\rho_{end}=\frac{3V_{end}}{2(1-X_{end})}\, ,
\end{equation}
where $X=\frac{8\kappa^2\dot\xi H}{\alpha}$. This is a more
simplified expression compared to \cite{ElBourakadi:2021nyb},
since the constraint $\ddot\xi=H\dot\xi$ has been imposed. In the
end, by combining all the above expressions, the duration of the
preheating era for the rescaled Einstein-Gauss-Bonnet model is
connected to the duration of reheating and inflation as follows,
\begin{align}
\centering \label{Npre}
N_{pre}&=61.6-\frac{1}{4}\ln\bigg(\frac{V_{end}}{\zeta
H_k^4(1-X_{end})}\bigg)-N_k-\ln
c_{\mathcal{S}}-\frac{1-3\omega_{pre}}{4}N_{re}\, ,
\end{align}
where assuming that $\beta\epsilon_1$ is small, one can expand the
sound wave velocity leaving us with $-\ln
c_\mathcal{S}\simeq2\beta^2\epsilon_1^2$ which serves as an small
or rather insignificant correction. It should also be stated that
by assuming that the preheating era is not necessary and after
inflation, reheating occurs, this suggests that $N_{pre}=0$ and
$\zeta=1$ leaving us with an expression for $N_{re}$ identical to
the one extracted in Ref. \cite{Venikoudis:2022gfg}.
Furthermore, for a vanishing Gauss-Bonnet coupling, it becomes
clear that $X_{end}=0$ in the above expression therefore the
result for the canonical scalar field for preheating \cite{ElBourakadi:2021blc} or reheating \cite{Cook:2015vqa}
emerges as it should. Obviously the special case of
$\omega_{pre}=\frac{1}{3}$ cannot be used here. Regarding the
expression for the duration of reheating, since
$N_{re}=\frac{1}{3(1+\omega_{pre})}\ln\frac{\rho_{pre}}{\rho_{re}}$,
by following similar steps it turns out that,
\begin{equation}
\centering \label{Nre}
N_{re}=\frac{1}{3(1+\omega_{pre})}\ln\bigg(\frac{45V_{end}}{\pi^2g_*\zeta
T_{re}^4(1-X_{end})}\bigg)\, ,
\end{equation}
with $T_{re}$ being a free parameter now and is specified by the
inflaton decay rate as \cite{Mazumdar:2013gya,Ellis:2015pla}
$T_{re}=\bigg(\frac{90}{8\pi^3g_*}\bigg)^{\frac{1}{4}}\sqrt{\Gamma_\phi
M_{Pl}}$. Note also that $\zeta$ shifts the value of $N_{re}$ and
given that a viable phenomenology requires $\zeta\gg 1$, the
duration of reheating decreases due to the fact that preheating
serves as an intermediate era. Hence, for the preheating era, one
requires the values for $\omega_{pre}$, $\zeta$ and $T_{re}$ in
order to extract numerical results, in contrast to the reheating
case, where either a value for $\omega_{re}\neq\frac{1}{3}$ or
both $N_{re}$ and $T_{re}$ are required. Hence, the preheating era
suggests an increase in the free parameters of the model by one,
and should be constrained from physical arguments. The value of
the reheating temperature is still unknown however a viable
reheating era can be obtain for values between $10^6-10^{12}$GeV,
see also Ref. \cite{Hasegawa:2019jsa} for MeV reheating
temperature. Concerning the value of the Hubble rate expansion
during the first horizon crossing, one can obtain information
about the power-spectra
$\mathcal{P}_{\mathcal{S}}=\mathcal{A}_{\mathcal{S}}\bigg(\frac{k}{k_*}\bigg)^{n_{\mathcal{S}}-1}$
which is equal to \cite{Hwang:2005hb},
\begin{align}
\centering \label{Ps}
\mathcal{P}_{\mathcal{S}}&=\frac{\kappa^2H^2}{8\pi^2\alpha
}\bigg[1-\epsilon_1+\frac{n_{\mathcal{S}}-1}{2}\bigg(\ln
2k|\eta|-2+\gamma_E\bigg)\bigg]^2\frac{4(1+\epsilon_4)^2(1-X)}{(4\epsilon_1(1-X)^2+3X^2)c_{\mathcal{S}}^{4-n_{\mathcal{S}}}}\,
,
\end{align}
therefore at the pivot scale, one can show that,
\begin{align}
\centering \label{Hk}
H_k&\simeq\frac{\pi}{\kappa}\bigg(\frac{c_{\mathcal{S}}^{\frac{4-n_{\mathcal{S}}}{2}}}{1+\epsilon_4}\bigg)\sqrt{8\alpha\mathcal{A}_{\mathcal{S}}\frac{4\epsilon_1(1-X)^2+3X^2}{4(1-X)}}\simeq\frac{\pi}{\kappa}\sqrt{8\alpha\mathcal{A}_{\mathcal{S}}\epsilon_1c_{\mathcal{S}}^3}\,
,
\end{align}
due to the fact that $\epsilon_i,X\ll 1$ while
$c_{\mathcal{S}}\simeq 1$ at the start of inflation. This was
already considered in Ref. \cite{Venikoudis:2022gfg} and one
can connect this result to the tensor-to-scalar ratio as
$8\epsilon_1c_{\mathcal{S}}^3=\frac{r}{2}$ for the constrained
rescaled Einstein-Gauss-Bonnet models. Note also that during the
first horizon crossing, $X=2\beta x$. Another interesting issue
that should be addressed here is the generation of gravitational
waves during the preheating era. For the Einstein-Gauss-Bonnet
model this has already been considered in Ref. \cite{ElBourakadi:2021nyb} where it was shown that the energy density is
proportional to the duration of preheating as,
\begin{figure}[h!]
\centering
\includegraphics[width=15pc]{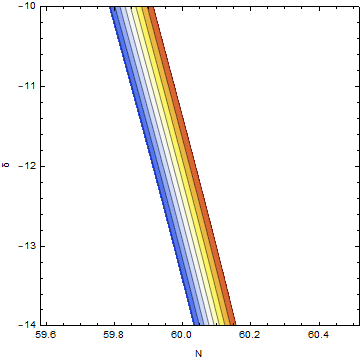}
\includegraphics[width=3pc]{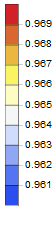}
\caption{Scalar spectral index as a function of $\delta$ and
e-folds $N$. As shown, the area of viability is quite slim for
these set of values. It should also be stated that $\lambda$ needs
to be fine tuned such that $n_{\mathcal{S}}$ deviates from unity.}
\label{fig1}
\end{figure}
\begin{equation}
\centering \label{preheatingGWs}
\Omega_{gw}(k)=\frac{\Omega_{gw,0}h^2}{\Omega_{r,0}h^2}\bigg(\frac{g_*}{g_0}\bigg)^{\frac{1}{3}}\e^{4N_{pre}}\,
,
\end{equation}
which in the context of Gauss-Bonnet model, it is mildly affected
by the factor $X=\frac{8\kappa^2\dot\xi H}{\alpha}$. In general,
the Gauss-Bonnet scalar coupling function and its dynamical
evolution according to Eq. (\ref{constraint1}) has interesting
applications in the energy spectrum of gravitational waves,
especially in the high frequency regime and we shall showcase this
explicitly in the following section. For the time being, let us
examine an inflationary model.
\begin{figure}[h!]
\centering
\includegraphics[width=15pc]{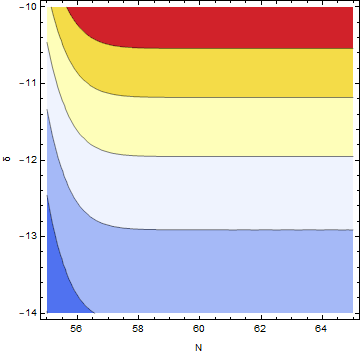}
\includegraphics[width=3pc]{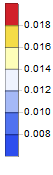}
\caption{Tensor-to-scalar ratio depending on $N$ and $\delta$. In
contrast to the scalar spectral index, the area of viability is
quite large and there exist a plethora of values capable of
satisfying the condition $r<0.064$.}
\label{fig2}
\end{figure}

\subsection{The case of $\xi(\phi)=\lambda\tanh(\delta\kappa\phi)\, $}

In this explicit example, we shall consider that the Gauss-Bonnet
scalar coupling function is known and is specified by the
following expression \cite{Rashidi:2020wwg},
\begin{equation}
\centering \label{xi} \xi(\phi)=\lambda\tanh(\delta\kappa\phi)\, ,
\end{equation}
where $\lambda$ and $\delta$ are arbitrary for the time being
dimensionless parameters with the first denoting the amplitude of
the coupling function. In order to specify the scalar potential,
we shall follow the steps mentioned before in the inflationary
phenomenology. Firstly, by solving the continuity equation under
the slow-roll assumption (\ref{motion6}) with respect to the
scalar potential, it becomes clear that for the aforementioned
Gauss-Bonnet coupling,
\begin{figure}[h!]
\centering
\includegraphics[width=15pc]{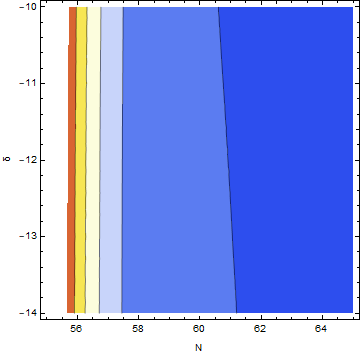}
\includegraphics[width=3pc]{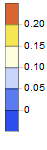}
\caption{Tensor spectral index $n_{\mathcal{T}}$ versus e-folds
and exponent $\delta$. As shown, there exist a lot of values that
manage to generate a blue spectrum.}
\label{fig3}
\end{figure}
\begin{widetext}
\begin{equation}
\centering \label{V} V(\phi)=\frac{1}{\Lambda\kappa^4  \sinh
^{-\frac{1}{2 \delta ^2}}(\delta  \kappa  \phi )-\frac{4}{3}
\kappa ^4 \lambda  \tanh ^2(\delta  \kappa  \phi
)^{\frac{1}{2}-\frac{1}{4 \delta ^2}} \coth (\delta  \kappa  \phi
) \text{sech}^2(\delta  \kappa  \phi )^{\frac{1}{4 \delta ^2}}
B_{\text{sech}^2(\delta  \kappa  \phi )}\left(1-\frac{1}{4 \delta
^2},\frac{1}{4} \left(2+\frac{1}{\delta ^2}\right)\right)}\, ,
\end{equation}
\end{widetext}
where $\Lambda$ is the integration parameter and $B_{\text{sech}}$
is the incomplete beta function. This is a quite non-trivial
potential, however it is completely specified by the free
parameters of the model. Note that hereafter for simplicity it is
assumed that $\alpha=1$. In consequence, since the inflationary
dynamics is mainly specified by the ratio
$\frac{\xi'}{\xi''}=-\frac{\coth (\delta  \kappa  \phi )}{2 \delta
\kappa }$, one can extract information about the scalar field. In
particular, with regards to the free parameters of the model, the
initial and final value of the scalar field during inflation
reads,
\begin{equation}
\centering \label{phik} \phi_k=\frac{1}{\delta\kappa}\cosh
^{-1}\left(\frac{e^{N/2}}{\sqrt{1-\frac{1}{8 \delta ^2}}}\right)\,
,
\end{equation}
\begin{equation}
\centering \label{phiend} \phi_{end}=-\frac{\coth ^{-1}\left(2
\sqrt{2}\delta \right)}{\delta  \kappa }\, .
\end{equation}
Let us see how the model could yield viable results. Assuming that
$(N,\delta,\lambda,\Lambda)=(60,-12,-10^{33},3\times10^{9})$, it
can easily be inferred that $n_{\mathcal{S}}=0.966794$,
$r=0.0138885$ and $n_{\mathcal{T}}=0.0025591$ which are in
agreement with observations \cite{Planck:2018vyg}, see Figs.
\ref{fig1}, \ref{fig2}, \ref{fig3} and see also Fig.
\ref{planck2018} for a more detailed analysis where the resulting
phenomenology is confronted with the latest Planck likelihood
curves.
\begin{figure}[h!]
\centering
\includegraphics[width=26pc]{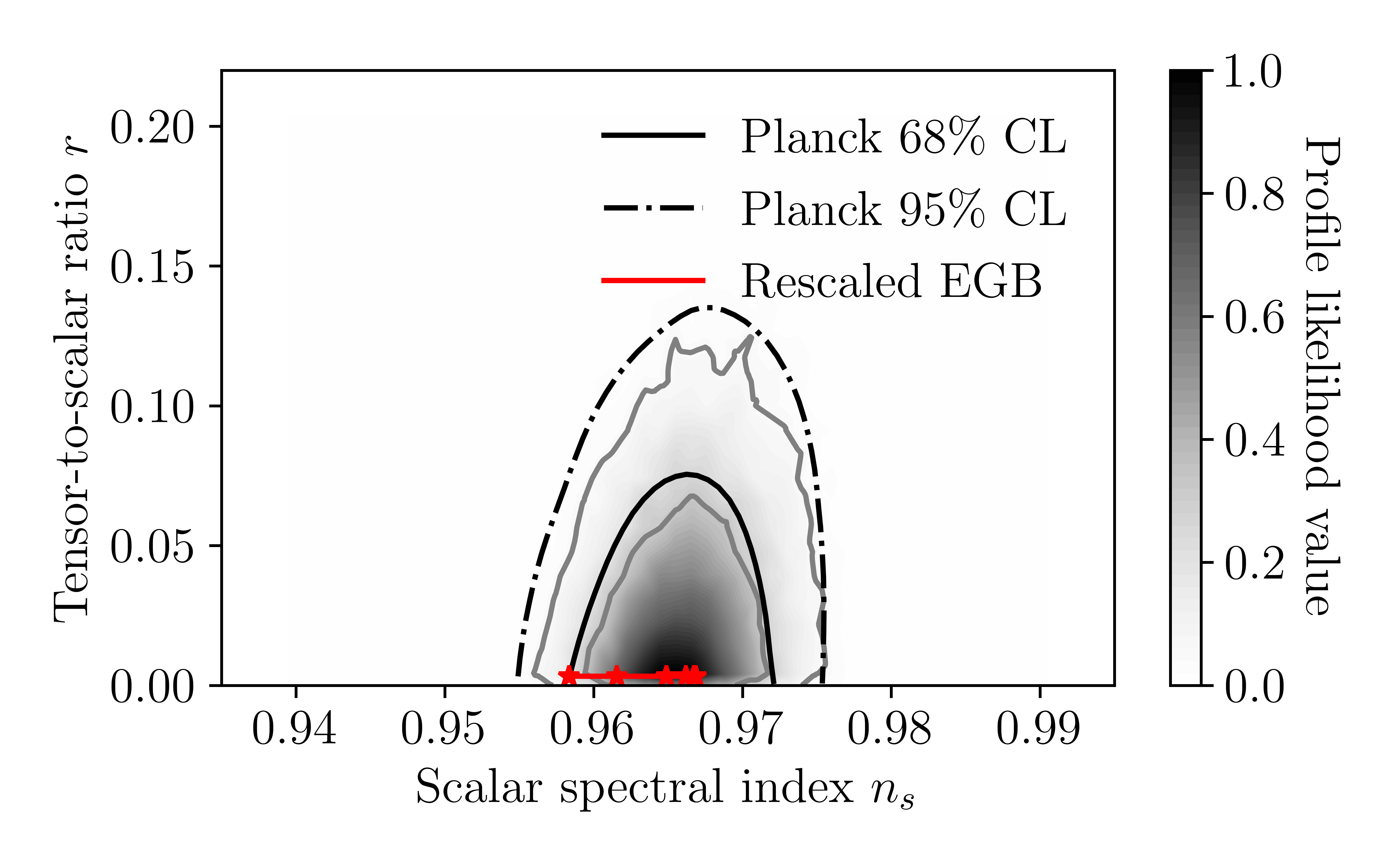}
\caption{The phenomenological confrontation of the model
$\xi(\phi)=\lambda\tanh(\delta\kappa\phi)\, $ with the latest
Planck likelihood curves.} \label{planck2018}
\end{figure}
It is interesting to mention that the tensor spectral index is
positive due to the fact that $\beta(\phi_k)=2.47491$ is greater
than unity and as it was proposed in the previous sections it is also worth stating that the Swampland criteria are
met since
$-\frac{V''(\phi_k)}{\kappa^2V(\phi_k)}=2.43902>\mathcal{O}(1)$. For this model, one
finds that $\phi_k=-2.5578M_{Pl}$ while
$\phi_{end}=-0.002456M_{Pl}$ which states that the scalar field
reaches zero as time flows by. The form of the scalar potential is
shown in Fig. \ref{fig4}. Furthermore, the slow-roll conditions
are met since $\epsilon_1=0.000868=-\epsilon_2$,
$\epsilon_3=0.01572$ and $\epsilon_4=-0.00215$.
\begin{figure}[h!]
\centering
\includegraphics[width=20pc]{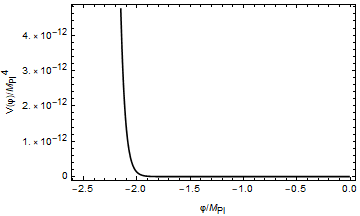}
\caption{Diagrammatic representation of the scalar potential. An
shown, negative values for the scalar field result in an apparent
increase of the strength of the potential while the limit $V\to0$
is reached for $\phi=0$ which serves as the minimum value. }
\label{fig4}
\end{figure}
Now in order to study subsequent cosmological eras, the value for
$X_{end}$ is required. This can be derived from the very
definition of $X=\frac{8\kappa^2\dot\xi H}{\alpha}$ at the end of
inflation where $\dot\phi_{end}^2=V_{end}$. In the end, one finds
that,
\begin{equation}
\centering \label{Xend}
X_{end}=\pm\frac{4\sqrt{2}\kappa^3V_{end}\xi'_{end}}{\alpha^{\frac{3}{2}}\sqrt{1-X_{end}}}\,
,
\end{equation}
which can be solved numerically for the same set of parameters
that were used in the inflationary phenomenology. Note that a
viable approach requires $X_{end}<1$, however there exists no
limit if $X<0$. Afterwards, by specifying $T_{re}$, $\omega_{pre}$
and substituting $\zeta$, information about the preheating and
preheating era can be extracted. This is shown in Figs. \ref{fig5}
and Fig. \ref{fig6} respectively, where as shown, while the
duration of preheating increases with $\omega_{pre}$, the duration
of the reheating era decreases. For a typical example, consider
the case of $\zeta=10^3$ with $\omega_{pre}=\frac{1}{5}$ and
$T_{re}=10^8$GeV where it turns out that $N_{pre}\simeq11.6$ and
$N_{re}\simeq0.81$.
\begin{figure}[h!]
\centering
\includegraphics[width=20pc]{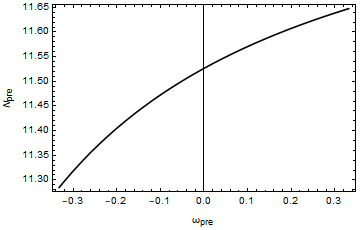}
\caption{Duration of preheating $N_{pre}$ as a function of the
preheating EoS. It becomes clear that the duration increases with
the increase of $\omega_{pre}$. For this numerical analysis, it is
assumed that $T_{re}=10^8$GeV and $\zeta=10^3$.}
\label{fig5}
\end{figure}

\section{Amplification of the energy spectrum of primordial gravitational waves}

Finally, in this section we shall briefly discuss the impact that
modifications in Einstein's gravity have on the energy spectrum of
gravitational waves. In order to extract information about the
impact that the Gauss-Bonnet scalar coupling function has on the
overall phenomenology, one needs to study the behavior of tensor
perturbations. Suppose that the perturbed metric now reads,
\begin{figure}[h!]
\centering
\includegraphics[width=20pc]{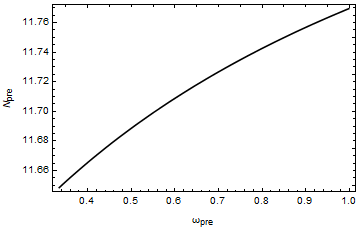}
\caption{Reheating duration for the same set of parameters. In
contrast to $N_{pre}$, $N_{re}$ seems to decrease with the
increase of the $\omega_{pre}$.}
\label{fig6}
\end{figure}
\begin{equation}
\centering \label{perturbedmetric}
ds^2=-dt^2+a^2(t)(\delta_{ij}+h_{ij}(t,\bm x))dx^idx^j\, ,
\end{equation}
where $h_{ij}(t,\bm x)$ describes tensor perturbations in
configuration space. For consistency, tensor perturbations need to
be traceless and transverse therefore the conditions
$\delta^{ij}h_{ij}=0=\partial_ih^{ij}$ must be respected. In
consequence, by varying the perturbed action, the Fourier
transformation of tensor modes must satisfy the following
equation,
\begin{equation}
\centering \label{modeeq} \ddot h(k)+(3+a_M)H\dot
h(k)+\frac{k^2}{a^2}h(k)=0\, ,
\end{equation}
where $a_M=\frac{\dot Q_t}{HQ_t}$ denotes the running of the
Planck mass \cite{Odintsov:2021kup,Odintsov:2022cbm}. Note that
the Gauss-Bonnet model is a subclass of Horndeski's theory that
affects the Planck mass therefore tensor modes are subsequently
affected. This was showcased previously when the tensor spectral
index was presented. Usually, for GR, $a_M=0$ therefore the
solution is known. Due to the presence of the running Planck mass,
one needs to extract a new solution for Eq. (\ref{modeeq}). This
can be achieved by making use of the WKB approximation
\cite{Odintsov:2021kup} where the solution is connected to the GR
result as shown below,
\begin{equation}
\centering \label{hsolution} h=\e^{-\mathcal{D}}h_{GR}\, ,
\end{equation}
where $h_{ij}=he_{ij}$ and
$\mathcal{D}=\int_0^zdz'\frac{a_M}{1+z'}$. This in turn implies
that modified theories of gravity which predict a nontrivial
running for the Planck mass manage, to affect the energy spectrum
of gravitational waves by means of the exponential factor
$\e^{-\mathcal{D}}$. This factor can result in an effected dumping
or amplification of the energy spectrum. This becomes clear from
the definition of the energy spectrum which in this case reads
\cite{Odintsov:2021kup},
\begin{equation}
\centering \label{energyspectrum}
\Omega_{gw}(k)=\frac{1}{\rho_{crit}}\frac{d\bra{0}\rho_{gw}\ket{0}}{d\ln
k}=\frac{k^2\Delta_h^2}{12H_0^2}\, ,
\end{equation}
where $\Delta_h^2=\frac{d\bra{0}h_{ij}^2\ket{0}}{d\ln k}$ is the
tensor power spectrum. In the context of modified theories of
gravity, the tensor power spectrum is given by the following
expression
\cite{Kamionkowski:2015yta,Turner:1993vb,Boyle:2005se,Zhang:2005nw,Caprini:2018mtu,Clarke:2020bil,Smith:2005mm,Giovannini:2008tm,Liu:2015psa,Vagnozzi:2020gtf,Kamionkowski:1993fg,Giare:2020vss,Zhao:2006mm,Lasky:2015lej,
Cai:2021uup,Benetti:2021uea,Lin:2021vwc,Zhang:2021vak,Pritchard:2004qp,Khoze:2022nyt,Oikonomou:2022xoq,ElBourakadi:2022anr,Arapoglu:2022vbf,Giare:2022wxq},
\begin{align}
\centering \label{tensorpowerspectrum}
\Delta_h^2&=\e^{-2\mathcal{D}}\Delta_h^{(p)2}\bigg(\frac{\Omega_m}{\Omega_\Lambda}\bigg)^2\bigg(\frac{g_{*}(T_{in})}{g_{*0}}\bigg)\bigg(\frac{g_{*s0}}{g_{*s}(T_{in})}\bigg)^{\frac{4}{3}}\bigg(\overline{\frac{3j_1(k\tau_0)}{k\tau_0}}\bigg)^2
T_1^2(x_{eq})T_2^2(x_R)\, ,
\end{align}
where $g_*$ denotes the relativistic degrees of freedom at
different cosmological eras, hence the reason why it is a
temperature dependent quantity, $T_i$ specifies the transfer
function at a given era dictates by $x$ witch denotes the ratio of
the pivot scale with either the matter-radiation equivalence of
the reheating scale and finally, $\Delta_h^{(p)2}$ denotes the
primordial tensor power spectrum which is connected to the
amplitude of tensor perturbations as,
\begin{equation}
\centering \label{primordialspectrum}
\Delta_h^{(p)2}=\mathcal{A}_{\mathcal{T}}\bigg(\frac{k}{k_*}\bigg)^{n_{\mathcal{T}}}\,
.
\end{equation}
At this point, it should be stated that the amplitude of tensor
perturbations, in contrast to scalar perturbations, has yet to be
observed however it is connected to the tensor-to-scalar ratio as
$\mathcal{A}_{\mathcal{T}}=r\mathcal{P}_\zeta$. Furthermore, the
tensor spectral index that participates in the exponent should be
treated as a frequency dependent parameter. In
the end, by using Eq. (\ref{running}) and by combining all the above expressions, one can show that
\cite{Odintsov:2021kup},
\begin{widetext}
\begin{equation}
\centering \label{energyspectrumfinal}
\Omega_{gw}(k)=\e^{-2\mathcal{D}}\frac{k^2}{12H_0^2}r\mathcal{P}_{\zeta}(k_*)\bigg(\frac{k}{k_*}\bigg)^{n_{\mathcal{T}}(k_*)+\frac{a_{\mathcal{T}(k_*)}}{2}\ln\frac{k}{k_*}}\bigg(\frac{\Omega_m}{\Omega_\Lambda}\bigg)^2\bigg(\frac{g_{*}(T_{in})}{g_{*0}}\bigg)\bigg(\frac{g_{*s0}}{g_{*s}(T_{in})}\bigg)^{\frac{4}{3}}\bigg(\overline{\frac{3j_1(k\tau_0)}{k\tau_0}}\bigg)^2
T_1^2(x_{eq})T_2^2(x_R)\, .
\end{equation}
\end{widetext}
At this point, two key features need to be highlighted. Firstly, a
negative factor $\mathcal{D}$ can in fact result in an apparent
enhancement of the energy spectrum measured today at all frequency
ranges. Secondly, provided that a blue-tilted tensor spectral
index is produced for $\beta>1$, it becomes clear that modes which
become subhorizon in primordial eras right after inflation, result
in an additional apparent amplification of the energy spectrum,
however this is possible only in the high frequency regime. This
is an interesting statement since these modes carry information
about subsequent eras such as reheating and radiation domination
and by simply studying high frequencies, information about such
eras can be obtained. Modified theories of gravity can in
principle result in an apparent enhancement or dumping thus making
the detection of gravitational waves at these frequencies either
an easier or more difficult task.

For the model at hand, the modified Planck mass has already been
computed in terms of the auxiliary parameter
$Q_t=\frac{\alpha}{\kappa^2}-8\dot\xi H$ denoting the shifted
Planck mass squared. In consequence, if one wants to compute the
running Planck mass $a_M$, provided that the constraint
$\ddot\xi=H\dot\xi$ is imposed, then it can easily be inferred
that,
\begin{equation}
\centering \label{aM1} a_M=-\frac{X}{1-X}\bigg[1+\frac{\dot
H}{H^2}\bigg]\, ,
\end{equation}
where $X=\frac{8\kappa^2\dot\xi H}{\alpha}$. Hence, the
deceleration parameter $q=-1-\frac{\dot H}{H^2}$ appears in the
running Planck mass. Now assuming that $1-X>0$, $X<0$, the running
Planck mass is negative during deceleration hence the amplitude
can in principle be enhanced. In general, by solving Eq.
(\ref{motion2}) in the presence of perfect matter fluids with
respect to $\dot H$, the deceleration parameter can be extracted
and in consequence, one can show that the running Planck mass can
be written with respect to dynamical variables as,
\begin{equation}
\centering \label{aM2}
a_M=\frac{X}{(1-X)^2}\bigg[-(1-X)+3Y^2+\frac{4\Omega_{\alpha
r}+3\Omega_{\alpha m}}{2}\bigg]\, ,
\end{equation}
where $Y=\frac{\kappa\dot\phi}{\sqrt{6\alpha}H}$ and
$\Omega_{\alpha i}=\frac{\kappa^2\rho_i}{3\alpha H^2}$. Generally
speaking, for $\alpha\to 1$ the usual description emerges however
under the assumption that $\alpha$ is an effective rescaled
version of the Einstein-Hilbert action due to the fact that higher
order curvature corrections provided by an $f(R)$ gravity result
in an effective shift of the coefficient of the Ricci scalar
primordially, one can extract information about the deceleration
parameter by studying $z=\frac{R}{6H^2}$ and in this approach,
\begin{equation}
\centering \label{aM3} a_M=\frac{X}{1-X}(1-z)\, .
\end{equation}
As it proves however, the amplification caused by the extra $a_M$
contribution on the energy spectrum of the primordial
gravitational waves is insignificant and of the order unity for
many rescaled Einstein-Gauss-Bonnet models we examined, this is
also common for ordinary Einstein-Gauss-Bonnet model
\cite{Oikonomou:2022xoq}. Thus the amplification of the primordial
gravitational waves energy spectrum can only occur due to a
significantly large and blue tilted tensor spectral index.

\section{Conclusions}

In the present article the inflationary dynamics of the
Einstein-Gauss-Bonnet theory for an arbitrary Gauss-Bonnet scalar
coupling functions was considered. It was shown that under the
assumption that tensor perturbations propagate through spacetime
with the velocity of light, therefore predicting primordial
massless gravitons and also being in agreement with the GW170817
event, the degrees of freedom still remain exactly the same as in
the case of a canonical scalar field. Afterwards, the spectral
indices and the tensor-to-scalar ratio were derived by making use
of several auxiliary dimensionless indices and as shown the
numerical value of the observables is completely dictated by three
parameters of the model, namely the first slow-roll index
$\epsilon_1$ which quantifies inflation, and parameters $\beta$
and $\gamma$ that carry information about the properties of the
Gauss-Bonnet scalar coupling function. In turn, the free
parameters of a given model should completely specify the above
three parameters. Afterwards the running of the spectral indices
was considered briefly were it was shown that regardless of the
fact that a constrained Einstein-Gauss-Bonnet model was studied,
which states that a handful of parameters may not necessarily be
small (compatible with a slow-roll evolution) during the first
horizon crossing, due to the fact that $\ddot\xi=H\dot\xi$, the
value of the first derivatives $a_{\mathcal{S}/\mathcal{T}}(k_*)$
evaluated at the pivot scale seems to be in agreement with
observations. Finally, the Swampland criteria were briefly
examined, were it was shown that while the second and third
criterion scale as $\sqrt{\epsilon_1}$ and $\epsilon_1$
respectively, it is possible to become simultaneously greater than
unity if a small rescaling parameter is used. In addition, having
$\beta>1$ seems to satisfy the third criterion while a blue-tilted
tensor spectral index is obtained. Also it was proved that the
tracking condition can be satisfied in the constrained
Gauss-Bonnet model only if the Swampland criteria are met
simultaneously, however a blue-tilted tensor spectral index cannot
be derived in that case. In addition, the model $\xi\sim 1/V$, along
with $\xi\sim V$, were excluded from the tracking condition since
the first is possible only for power-law model with an exponent
lesser than unity while the latter is possible for exponential
scalar functions which cannot describe the inflationary era
properly in the constrained Gauss-Bonnet formalism. Afterwards
the preheating era was briefly discussed where the duration $N_{pre}$
was derived as a function of the duration of the reheating era and
temperature as well as the effective EoS. Assuming that the propagation
velocity of tensor perturbations is fixed to coincide with the speed of
light, it was shown that the energy density at the end of inflation is
shifted by a factor of $1/1-X$ where $X=\frac{8\kappa^2\dot\xi H}{\alpha}$.
The model was also tested by assuming that the Gauss-Bonnet scalar coupling
function has a hyperbolic form $\xi(\phi)=\lambda\tanh(\delta\kappa\phi)$ where
it was shown that the model can indeed be in agreement with observations.
In consequence, a brief comment on the energy spectrum of primordial gravitational
waves was made and as shown, the energy spectrum in the constrained Gauss-Bonnet
model can be enhanced by either a negative factor $\mathcal{D}$ or by a
positive value of the tensor spectral index, or in other words for $\beta>1$.
The latter case is connected to the amplification of the energy spectrum of
primordial gravitational waves in the high frequency region, in which modes
re-enter the horizon relatively fast and can thus shine light towards
primordial cosmological eras of interest. From a theoretical perspective,
it would be interesting to see how certain inflationary models can yield
viable results under the assumption that $\ddot\xi=H\dot\xi$ and
also whether the Swampland criteria, if satisfied, are in
agreement with the tracking condition. Furthermore, the
possibility of a kinetic coupling of the form
$\frac{1}{2}\xi(\phi)G^{\mu\nu}\nabla_\mu\phi\nabla_\nu\phi$ in
the gravitational action may be promising since the constraint on
the propagation velocity of tensor perturbations is now altered.
We aim to address these interesting topics, along with a complete
study of the rescaled Einstein-Gauss-Bonnet theories
phenomenology, in future works.

\section*{Acknowledgments}

This work was supported by MINECO (Spain), project
PID2019-104397GB-I00 (S.D.O). This work by S.D.O was also
partially supported by the program Unidad de Excelencia Maria de
Maeztu CEX2020-001058-M, Spain.


\begin{thebibliography}{99}




\bibitem{inflation1}
 A.~D.~Linde,
 Lect.\ Notes Phys.\ {\bf 738} (2008) 1
 [arXiv:0705.0164 [hep-th]].

\bibitem{inflation2} D.~S.~Gorbunov and V.~A.~Rubakov,
``Introduction to the theory of the early universe: Cosmological
perturbations and inflationary theory,'' Hackensack, USA: World
Scientific (2011) 489 p;
%


\bibitem{inflation3}A.~Linde,
arXiv:1402.0526 [hep-th];


\bibitem{inflation4}D.~H.~Lyth and A.~Riotto,
Phys.\ Rept.\  {\bf 314} (1999) 1 [hep-ph/9807278].






\bibitem{Baker:2019nia}
J.~Baker, J.~Bellovary, P.~L.~Bender, E.~Berti, R.~Caldwell,
J.~Camp, J.~W.~Conklin, N.~Cornish, C.~Cutler and R.~DeRosa,
\textit{et al.}
[arXiv:1907.06482 [astro-ph.IM]].


\bibitem{Smith:2019wny}
T.~L.~Smith and R.~Caldwell,
Phys. Rev. D \textbf{100} (2019) no.10, 104055
doi:10.1103/PhysRevD.100.104055 [arXiv:1908.00546 [astro-ph.CO]].


\bibitem{Crowder:2005nr}
J.~Crowder and N.~J.~Cornish,
Phys. Rev. D \textbf{72} (2005), 083005
doi:10.1103/PhysRevD.72.083005 [arXiv:gr-qc/0506015 [gr-qc]].


\bibitem{Smith:2016jqs}
T.~L.~Smith and R.~Caldwell,
Phys. Rev. D \textbf{95} (2017) no.4, 044036
doi:10.1103/PhysRevD.95.044036 [arXiv:1609.05901 [gr-qc]].



\bibitem{Seto:2001qf}
N.~Seto, S.~Kawamura and T.~Nakamura,
Phys. Rev. Lett. \textbf{87} (2001), 221103
doi:10.1103/PhysRevLett.87.221103 [arXiv:astro-ph/0108011
[astro-ph]].


\bibitem{Kawamura:2020pcg}
S.~Kawamura, M.~Ando, N.~Seto, S.~Sato, M.~Musha, I.~Kawano,
J.~Yokoyama, T.~Tanaka, K.~Ioka and T.~Akutsu, \textit{et al.}
[arXiv:2006.13545 [gr-qc]].




\bibitem{CMB-S4:2016ple}
K.~N.~Abazajian \textit{et al.} [CMB-S4],
[arXiv:1610.02743 [astro-ph.CO]].



\bibitem{SimonsObservatory:2019qwx}
M.~H.~Abitbol \textit{et al.} [Simons Observatory],
Bull. Am. Astron. Soc. \textbf{51} (2019), 147 [arXiv:1907.08284
[astro-ph.IM]].



\bibitem{Hwang:2005hb}
  J.~c.~Hwang and H.~Noh,
  Phys.\ Rev.\ D {\bf 71} (2005) 063536
  doi:10.1103/PhysRevD.71.063536
  [gr-qc/0412126].


\bibitem{Nojiri:2006je}
  S.~Nojiri, S.~D.~Odintsov and M.~Sami,
  Phys.\ Rev.\ D {\bf 74} (2006) 046004
  doi:10.1103/PhysRevD.74.046004
  [hep-th/0605039].




\bibitem{Cognola:2006sp}
  G.~Cognola, E.~Elizalde, S.~Nojiri, S.~Odintsov and S.~Zerbini,
  Phys.\ Rev.\ D {\bf 75} (2007) 086002
  doi:10.1103/PhysRevD.75.086002
  [hep-th/0611198].



\bibitem{Nojiri:2005vv}
  S.~Nojiri, S.~D.~Odintsov and M.~Sasaki,
  Phys.\ Rev.\ D {\bf 71} (2005) 123509
  doi:10.1103/PhysRevD.71.123509
  [hep-th/0504052].


\bibitem{Nojiri:2005jg}
  S.~Nojiri and S.~D.~Odintsov,
  Phys.\ Lett.\ B {\bf 631} (2005) 1
  doi:10.1016/j.physletb.2005.10.010
  [hep-th/0508049].







\bibitem{Satoh:2007gn}
  M.~Satoh, S.~Kanno and J.~Soda,
  Phys.\ Rev.\ D {\bf 77} (2008) 023526
  doi:10.1103/PhysRevD.77.023526
  [arXiv:0706.3585 [astro-ph]].



\bibitem{Bamba:2014zoa}
  K.~Bamba, A.~N.~Makarenko, A.~N.~Myagky and S.~D.~Odintsov,
  JCAP {\bf 1504} (2015) 001
  doi:10.1088/1475-7516/2015/04/001
  [arXiv:1411.3852 [hep-th]].


\bibitem{Yi:2018gse}
  Z.~Yi, Y.~Gong and M.~Sabir,
  Phys.\ Rev.\ D {\bf 98} (2018) no.8,  083521
  doi:10.1103/PhysRevD.98.083521
  [arXiv:1804.09116 [gr-qc]].


\bibitem{Guo:2009uk}
  Z.~K.~Guo and D.~J.~Schwarz,
  Phys.\ Rev.\ D {\bf 80} (2009) 063523
  doi:10.1103/PhysRevD.80.063523
  [arXiv:0907.0427 [hep-th]].


\bibitem{Guo:2010jr}
  Z.~K.~Guo and D.~J.~Schwarz,
  Phys.\ Rev.\ D {\bf 81} (2010) 123520
  doi:10.1103/PhysRevD.81.123520
  [arXiv:1001.1897 [hep-th]].


\bibitem{Jiang:2013gza}
  P.~X.~Jiang, J.~W.~Hu and Z.~K.~Guo,
  Phys.\ Rev.\ D {\bf 88} (2013) 123508
  doi:10.1103/PhysRevD.88.123508
  [arXiv:1310.5579 [hep-th]].



\bibitem{Kanti:2015pda}
  P.~Kanti, R.~Gannouji and N.~Dadhich,
  Phys.\ Rev.\ D {\bf 92} (2015) no.4,  041302
  doi:10.1103/PhysRevD.92.041302
  [arXiv:1503.01579 [hep-th]].


\bibitem{vandeBruck:2017voa}
  C.~van de Bruck, K.~Dimopoulos, C.~Longden and C.~Owen,
  arXiv:1707.06839 [astro-ph.CO].



\bibitem{Kanti:1998jd}
  P.~Kanti, J.~Rizos and K.~Tamvakis,
  Phys.\ Rev.\ D {\bf 59} (1999) 083512
  doi:10.1103/PhysRevD.59.083512
  [gr-qc/9806085].



\bibitem{Pozdeeva:2020apf}
E.~O.~Pozdeeva, M.~R.~Gangopadhyay, M.~Sami, A.~V.~Toporensky and
S.~Y.~Vernov,
Phys. Rev. D \textbf{102} (2020) no.4, 043525
doi:10.1103/PhysRevD.102.043525 [arXiv:2006.08027 [gr-qc]].



\bibitem{Vernov:2021hxo}
S.~Vernov and E.~Pozdeeva,
Universe \textbf{7} (2021) no.5, 149 doi:10.3390/universe7050149
[arXiv:2104.11111 [gr-qc]].


\bibitem{Pozdeeva:2021iwc}
E.~O.~Pozdeeva and S.~Y.~Vernov,
Eur. Phys. J. C \textbf{81} (2021) no.7, 633
doi:10.1140/epjc/s10052-021-09435-8 [arXiv:2104.04995 [gr-qc]].




\bibitem{Koh:2014bka}
S.~Koh, B.~H.~Lee, W.~Lee and G.~Tumurtushaa,
Phys. Rev. D \textbf{90} (2014) no.6, 063527
doi:10.1103/PhysRevD.90.063527 [arXiv:1404.6096 [gr-qc]].



\bibitem{Bayarsaikhan:2020jww}
B.~Bayarsaikhan, S.~Koh, E.~Tsedenbaljir and G.~Tumurtushaa,
JCAP \textbf{11} (2020), 057 doi:10.1088/1475-7516/2020/11/057
[arXiv:2005.11171 [gr-qc]].



\bibitem{Tumurtushaa:2018lnv}
G.~Tumurtushaa, S.~Koh and B.~H.~Lee,
PoS \textbf{ICHEP2018} (2019), 090 doi:10.22323/1.340.0090






\bibitem{Fomin:2020hfh}
I.~Fomin,
Eur. Phys. J. C \textbf{80} (2020) no.12, 1145
doi:10.1140/epjc/s10052-020-08718-w [arXiv:2004.08065 [gr-qc]].

\bibitem{DeLaurentis:2015fea}
  M.~De Laurentis, M.~Paolella and S.~Capozziello,
  Phys.\ Rev.\ D {\bf 91} (2015) no.8,  083531
  doi:10.1103/PhysRevD.91.083531
  [arXiv:1503.04659 [gr-qc]].


\bibitem{Chervon:2019sey}
   Scalar Field Cosmology, S.~Chervon, I.~Fomin, V.~Yurov and
   A.~Yurov, World Scientific 2019, \\  doi:10.1142/11405



\bibitem{Nozari:2017rta}
  K.~Nozari and N.~Rashidi,
  Phys.\ Rev.\ D {\bf 95} (2017) no.12,  123518
  doi:10.1103/PhysRevD.95.123518
  [arXiv:1705.02617 [astro-ph.CO]].




\bibitem{Odintsov:2018zhw}
  S.~D.~Odintsov and V.~K.~Oikonomou,
  Phys.\ Rev.\ D {\bf 98} (2018) no.4,  044039
  doi:10.1103/PhysRevD.98.044039
  [arXiv:1808.05045 [gr-qc]].


  \bibitem{Kawai:1998ab}
  S.~Kawai, M.~a.~Sakagami and J.~Soda,
  Phys.\ Lett.\ B {\bf 437}, 284 (1998)
  doi:10.1016/S0370-2693(98)00925-3
  [gr-qc/9802033].


\bibitem{Yi:2018dhl}
  Z.~Yi and Y.~Gong,
  Universe {\bf 5} (2019) no.9,  200
  doi:10.3390/universe5090200
  [arXiv:1811.01625 [gr-qc]].


\bibitem{vandeBruck:2016xvt}
  C.~van de Bruck, K.~Dimopoulos and C.~Longden,
  Phys.\ Rev.\ D {\bf 94} (2016) no.2,  023506
  doi:10.1103/PhysRevD.94.023506
  [arXiv:1605.06350 [astro-ph.CO]].


\bibitem{Kleihaus:2019rbg}
B.~Kleihaus, J.~Kunz and P.~Kanti,
Phys. Lett. B \textbf{804} (2020), 135401
doi:10.1016/j.physletb.2020.135401 [arXiv:1910.02121 [gr-qc]].




\bibitem{Bakopoulos:2019tvc}
  A.~Bakopoulos, P.~Kanti and N.~Pappas,
  Phys.\ Rev.\ D {\bf 101} (2020) no.4,  044026
  doi:10.1103/PhysRevD.101.044026
  [arXiv:1910.14637 [hep-th]].


\bibitem{Maeda:2011zn}
  K.~i.~Maeda, N.~Ohta and R.~Wakebe,
  Eur.\ Phys.\ J.\ C {\bf 72} (2012) 1949
  doi:10.1140/epjc/s10052-012-1949-6
  [arXiv:1111.3251 [hep-th]].





\bibitem{Bakopoulos:2020dfg}
A.~Bakopoulos, P.~Kanti and N.~Pappas,
Phys. Rev. D \textbf{101} (2020) no.8, 084059
doi:10.1103/PhysRevD.101.084059 [arXiv:2003.02473 [hep-th]].

\bibitem{Ai:2020peo}
W.~Y.~Ai,
Commun. Theor. Phys. \textbf{72} (2020) no.9, 095402
doi:10.1088/1572-9494/aba242 [arXiv:2004.02858 [gr-qc]].



\bibitem{Oikonomou:2020oil}
V.~K.~Oikonomou and F.~P.~Fronimos,
EPL \textbf{131} (2020) no.3, 30001
doi:10.1209/0295-5075/131/30001 [arXiv:2007.11915 [gr-qc]].

\bibitem{Odintsov:2020xji}
S.~D.~Odintsov, V.~K.~Oikonomou and F.~P.~Fronimos,
Annals Phys. \textbf{420} (2020), 168250
doi:10.1016/j.aop.2020.168250 [arXiv:2007.02309 [gr-qc]].



\bibitem{Oikonomou:2020sij}
V.~K.~Oikonomou and F.~P.~Fronimos,
Class. Quant. Grav. \textbf{38} (2021) no.3, 035013
doi:10.1088/1361-6382/abce47 [arXiv:2006.05512 [gr-qc]].



\bibitem{Odintsov:2020zkl}
S.~D.~Odintsov and V.~K.~Oikonomou,
Phys. Lett. B \textbf{805} (2020), 135437
doi:10.1016/j.physletb.2020.135437 [arXiv:2004.00479 [gr-qc]].




\bibitem{Odintsov:2020mkz}
S.~D.~Odintsov, V.~K.~Oikonomou, F.~P.~Fronimos and
S.~A.~Venikoudis,
Phys. Dark Univ. \textbf{30} (2020), 100718
doi:10.1016/j.dark.2020.100718 [arXiv:2009.06113 [gr-qc]].


\bibitem{Venikoudis:2021irr}
S.~A.~Venikoudis and F.~P.~Fronimos,
Eur. Phys. J. Plus \textbf{136} (2021) no.3, 308
doi:10.1140/epjp/s13360-021-01298-y [arXiv:2103.01875 [gr-qc]].



\bibitem{Kong:2021qiu}
S.~B.~Kong, H.~Abdusattar, Y.~Yin and Y.~P.~Hu,
[arXiv:2108.09411 [gr-qc]].





\bibitem{Easther:1996yd}
  R.~Easther and K.~i.~Maeda,
  Phys.\ Rev.\ D {\bf 54} (1996) 7252
  doi:10.1103/PhysRevD.54.7252
  [hep-th/9605173].

\bibitem{Antoniadis:1993jc}
  I.~Antoniadis, J.~Rizos and K.~Tamvakis,
  Nucl.\ Phys.\ B {\bf 415} (1994) 497
  doi:10.1016/0550-3213(94)90120-1
  [hep-th/9305025].

\bibitem{Antoniadis:1990uu}
I.~Antoniadis, C.~Bachas, J.~R.~Ellis and D.~V.~Nanopoulos,
Phys.\ Lett.\ B \textbf{257} (1991), 278-284
doi:10.1016/0370-2693(91)91893-Z




\bibitem{Kanti:1995vq}
P.~Kanti, N.~Mavromatos, J.~Rizos, K.~Tamvakis and E.~Winstanley,
Phys. Rev. D \textbf{54} (1996), 5049-5058
doi:10.1103/PhysRevD.54.5049 [arXiv:hep-th/9511071 [hep-th]].



\bibitem{Kanti:1997br}
P.~Kanti, N.~Mavromatos, J.~Rizos, K.~Tamvakis and E.~Winstanley,
Phys. Rev. D \textbf{57} (1998), 6255-6264
doi:10.1103/PhysRevD.57.6255 [arXiv:hep-th/9703192 [hep-th]].



\bibitem{Easson:2020mpq}
D.~A.~Easson, T.~Manton and A.~Svesko,
JCAP \textbf{10} (2020), 026 doi:10.1088/1475-7516/2020/10/026
[arXiv:2005.12292 [hep-th]].


\bibitem{Rashidi:2020wwg}
N.~Rashidi and K.~Nozari,
Astrophys. J. \textbf{890}, 58
doi:10.3847/1538-4357/ab6a10
[arXiv:2001.07012 [astro-ph.CO]].









\bibitem{Oikonomou:2022xoq}
V.~K.~Oikonomou,
Astropart. Phys. \textbf{141} (2022), 102718
doi:10.1016/j.astropartphys.2022.102718 [arXiv:2204.06304
[gr-qc]].



\bibitem{TheLIGOScientific:2017qsa}
B.~P.~Abbott \textit{et al.} [LIGO Scientific and Virgo],
Phys. Rev. Lett. \textbf{119} (2017) no.16, 161101
doi:10.1103/PhysRevLett.119.161101 [arXiv:1710.05832 [gr-qc]].




\bibitem{Monitor:2017mdv}
B.~P.~Abbott \textit{et al.} [LIGO Scientific, Virgo, Fermi-GBM
and INTEGRAL],
Astrophys. J. Lett. \textbf{848} (2017) no.2, L13
doi:10.3847/2041-8213/aa920c [arXiv:1710.05834 [astro-ph.HE]].


\bibitem{GBM:2017lvd}
  B.~P.~Abbott {\it et al.}
  ``Multi-messenger Observations of a Binary Neutron Star Merger,''
  Astrophys.\ J.\  {\bf 848} (2017) no.2,  L12
  doi:10.3847/2041-8213/aa91c9
  [arXiv:1710.05833 [astro-ph.HE]].



\bibitem{Ezquiaga:2017ekz}
J.~M.~Ezquiaga and M.~Zumalac\'arregui,
Phys. Rev. Lett. \textbf{119} (2017) no.25, 251304
doi:10.1103/PhysRevLett.119.251304 [arXiv:1710.05901
[astro-ph.CO]].


\bibitem{Baker:2017hug}
T.~Baker, E.~Bellini, P.~G.~Ferreira, M.~Lagos, J.~Noller and
I.~Sawicki,
Phys. Rev. Lett. \textbf{119} (2017) no.25, 251301
doi:10.1103/PhysRevLett.119.251301 [arXiv:1710.06394
[astro-ph.CO]].


\bibitem{Creminelli:2017sry}
P.~Creminelli and F.~Vernizzi,
Phys. Rev. Lett. \textbf{119} (2017) no.25, 251302
doi:10.1103/PhysRevLett.119.251302 [arXiv:1710.05877
[astro-ph.CO]].


\bibitem{Sakstein:2017xjx}
J.~Sakstein and B.~Jain,
Phys. Rev. Lett. \textbf{119} (2017) no.25, 251303
doi:10.1103/PhysRevLett.119.251303 [arXiv:1710.05893
[astro-ph.CO]].


\bibitem{Odintsov:2020sqy}
S.~D.~Odintsov, V.~K.~Oikonomou and F.~P.~Fronimos,
Nucl. Phys. B \textbf{958} (2020), 115135
doi:10.1016/j.nuclphysb.2020.115135 [arXiv:2003.13724 [gr-qc]].


\bibitem{Oikonomou:2021kql}
V.~K.~Oikonomou,
Class. Quant. Grav. \textbf{38} (2021) no.19, 195025
doi:10.1088/1361-6382/ac2168 [arXiv:2108.10460 [gr-qc]].






\bibitem{Oikonomou:2022ksx}
V.~K.~Oikonomou, P.~D.~Katzanis and I.~C.~Papadimitriou,
Class. Quant. Grav. \textbf{39} (2022) no.9, 095008
doi:10.1088/1361-6382/ac5eba [arXiv:2203.09867 [gr-qc]].



















\bibitem{Vafa:2005ui}
C.~Vafa,
[arXiv:hep-th/0509212 [hep-th]].



\bibitem{Ooguri:2006in}
  H.~Ooguri and C.~Vafa,
  Nucl.\ Phys.\ B {\bf 766} (2007) 21
  doi:10.1016/j.nuclphysb.2006.10.033
  [hep-th/0605264].



\bibitem{Palti:2020qlc}
  E.~Palti, C.~Vafa and T.~Weigand,
  arXiv:2003.10452 [hep-th].



\bibitem{Brandenberger:2020oav}
  R.~Brandenberger, V.~Kamali and R.~O.~Ramos,
  arXiv:2002.04925 [hep-th].



\bibitem{Blumenhagen:2019vgj}
  R.~Blumenhagen, M.~Brinkmann and A.~Makridou,
  JHEP {\bf 2002} (2020) 064
   [JHEP {\bf 2020} (2020) 064]
  doi:10.1007/JHEP02(2020)064
  [arXiv:1910.10185 [hep-th]].



\bibitem{Wang:2019eym}
  Z.~Wang, R.~Brandenberger and L.~Heisenberg,
  arXiv:1907.08943 [hep-th].





\bibitem{Benetti:2019smr}
  M.~Benetti, S.~Capozziello and L.~L.~Graef,
  Phys.\ Rev.\ D {\bf 100} (2019) no.8,  084013
  doi:10.1103/PhysRevD.100.084013
  [arXiv:1905.05654 [gr-qc]].




\bibitem{Palti:2019pca}
  E.~Palti,
  Fortsch.\ Phys.\  {\bf 67} (2019) no.6,  1900037
  doi:10.1002/prop.201900037
  [arXiv:1903.06239 [hep-th]].




\bibitem{Cai:2018ebs}
  R.~G.~Cai, S.~Khimphun, B.~H.~Lee, S.~Sun, G.~Tumurtushaa and Y.~L.~Zhang,
  Phys.\ Dark Univ.\  {\bf 26} (2019) 100387
  doi:10.1016/j.dark.2019.100387
  [arXiv:1812.11105 [hep-th]].



\bibitem{Akrami:2018ylq}
Y.~Akrami, R.~Kallosh, A.~Linde and V.~Vardanyan,
Fortsch. Phys. \textbf{67} (2019) no.1-2, 1800075
doi:10.1002/prop.201800075 [arXiv:1808.09440 [hep-th]].

\bibitem{Mizuno:2019pcm}
  S.~Mizuno, S.~Mukohyama, S.~Pi and Y.~L.~Zhang,
  JCAP {\bf 1909} (2019) no.09,  072
  doi:10.1088/1475-7516/2019/09/072
  [arXiv:1905.10950 [hep-th]].




\bibitem{Aragam:2019khr}
  V.~Aragam, S.~Paban and R.~Rosati,
  arXiv:1905.07495 [hep-th].





\bibitem{Brahma:2019mdd}
  S.~Brahma and M.~W.~Hossain,
  Phys.\ Rev.\ D {\bf 100} (2019) no.8,  086017
  doi:10.1103/PhysRevD.100.086017
  [arXiv:1904.05810 [hep-th]].




\bibitem{Mukhopadhyay:2019cai}
  U.~Mukhopadhyay and D.~Majumdar,
  Phys.\ Rev.\ D {\bf 100} (2019) no.2,  024006
  doi:10.1103/PhysRevD.100.024006
  [arXiv:1904.01455 [gr-qc]].







\bibitem{Brahma:2019kch}
  S.~Brahma and M.~W.~Hossain,
  JHEP {\bf 1906} (2019) 070
  doi:10.1007/JHEP06(2019)070
  [arXiv:1902.11014 [hep-th]].



\bibitem{Haque:2019prw}
  M.~R.~Haque and D.~Maity,
  Phys.\ Rev.\ D {\bf 99} (2019) no.10,  103534
  doi:10.1103/PhysRevD.99.103534
  [arXiv:1902.09491 [hep-th]].

\bibitem{Heckman:2019dsj}
  J.~J.~Heckman, C.~Lawrie, L.~Lin, J.~Sakstein and G.~Zoccarato,
  Fortsch.\ Phys.\  {\bf 67} (2019) no.11,  1900071
  doi:10.1002/prop.201900071
  [arXiv:1901.10489 [hep-th]].

\bibitem{Acharya:2018deu}
  B.~S.~Acharya, A.~Maharana and F.~Muia,
  JHEP {\bf 1903} (2019) 048
  doi:10.1007/JHEP03(2019)048
  [arXiv:1811.10633 [hep-th]].






\bibitem{Elizalde:2018dvw}
E.~Elizalde and M.~Khurshudyan,
Phys.\ Rev.\ D \textbf{99} (2019) no.10, 103533
doi:10.1103/PhysRevD.99.103533 [arXiv:1811.03861 [astro-ph.CO]].


\bibitem{Cheong:2018udx}
  D.~Y.~Cheong, S.~M.~Lee and S.~C.~Park,
  Phys.\ Lett.\ B {\bf 789} (2019) 336
  doi:10.1016/j.physletb.2018.12.046
  [arXiv:1811.03622 [hep-ph]].

\bibitem{Heckman:2018mxl}
  J.~J.~Heckman, C.~Lawrie, L.~Lin and G.~Zoccarato,
  Fortsch.\ Phys.\  {\bf 67} (2019) no.10,  1900057
  doi:10.1002/prop.201900057
  [arXiv:1811.01959 [hep-th]].


\bibitem{Kinney:2018nny}
W.~H.~Kinney, S.~Vagnozzi and L.~Visinelli,
Class.\ Quant.\ Grav.\  \textbf{36} (2019) no.11, 117001
doi:10.1088/1361-6382/ab1d87 [arXiv:1808.06424 [astro-ph.CO]].



\bibitem{Garg:2018reu}
  S.~K.~Garg and C.~Krishnan,
  JHEP {\bf 1911} (2019) 075
  doi:10.1007/JHEP11(2019)075
  [arXiv:1807.05193 [hep-th]].

\bibitem{Lin:2018rnx}
  C.~M.~Lin,
  Phys.\ Rev.\ D {\bf 99} (2019) no.2,  023519
  doi:10.1103/PhysRevD.99.023519
  [arXiv:1810.11992 [astro-ph.CO]].

\bibitem{Park:2018fuj}
  S.~C.~Park,
  JCAP {\bf 1901} (2019) 053
  doi:10.1088/1475-7516/2019/01/053
  [arXiv:1810.11279 [hep-ph]].

\bibitem{Olguin-Tejo:2018pfq}
  Y.~Olguin-Trejo, S.~L.~Parameswaran, G.~Tasinato and I.~Zavala,
  JCAP {\bf 1901} (2019) 031
  doi:10.1088/1475-7516/2019/01/031
  [arXiv:1810.08634 [hep-th]].





\bibitem{Fukuda:2018haz}
  H.~Fukuda, R.~Saito, S.~Shirai and M.~Yamazaki,
  Phys.\ Rev.\ D {\bf 99} (2019) no.8,  083520
  doi:10.1103/PhysRevD.99.083520
  [arXiv:1810.06532 [hep-th]].


\bibitem{Wang:2018kly}
  S.~J.~Wang,
  Phys.\ Rev.\ D {\bf 99} (2019) no.2,  023529
  doi:10.1103/PhysRevD.99.023529
  [arXiv:1810.06445 [hep-th]].


\bibitem{Ooguri:2018wrx}
  H.~Ooguri, E.~Palti, G.~Shiu and C.~Vafa,
  Phys.\ Lett.\ B {\bf 788} (2019) 180
  doi:10.1016/j.physletb.2018.11.018
  [arXiv:1810.05506 [hep-th]].

\bibitem{Matsui:2018xwa}
  H.~Matsui, F.~Takahashi and M.~Yamada,
  Phys.\ Lett.\ B {\bf 789} (2019) 387
  doi:10.1016/j.physletb.2018.12.055
  [arXiv:1809.07286 [astro-ph.CO]].



\bibitem{Obied:2018sgi}
  G.~Obied, H.~Ooguri, L.~Spodyneiko and C.~Vafa,
  arXiv:1806.08362 [hep-th].


\bibitem{Agrawal:2018own}
  P.~Agrawal, G.~Obied, P.~J.~Steinhardt and C.~Vafa,
  Phys.\ Lett.\ B {\bf 784} (2018) 271
  doi:10.1016/j.physletb.2018.07.040
  [arXiv:1806.09718 [hep-th]].


\bibitem{Murayama:2018lie}
  H.~Murayama, M.~Yamazaki and T.~T.~Yanagida,
  JHEP {\bf 1812} (2018) 032
  doi:10.1007/JHEP12(2018)032
  [arXiv:1809.00478 [hep-th]].

\bibitem{Marsh:2018kub}
  M.~C.~David Marsh,
  Phys.\ Lett.\ B {\bf 789} (2019) 639
  doi:10.1016/j.physletb.2018.11.001
  [arXiv:1809.00726 [hep-th]].




\bibitem{Storm:2020gtv}
S.~D.~Storm and R.~J.~Scherrer,
Phys. Rev. D \textbf{102} (2020) no.6, 063519
doi:10.1103/PhysRevD.102.063519 [arXiv:2008.05465 [hep-th]].



\bibitem{Trivedi:2020wxf}
O.~Trivedi,
[arXiv:2008.05474 [hep-th]].



\bibitem{Sharma:2020wba}
U.~K.~Sharma,
[arXiv:2005.03979 [physics.gen-ph]].



\bibitem{Mohammadi:2020twg}
A.~Mohammadi, T.~Golanbari, J.~Enayati, S.~Jalalzadeh and
K.~Saaidi,
[arXiv:2011.13957 [gr-qc]].


\bibitem{Trivedi:2020xlh}
O.~Trivedi,
[arXiv:2011.14316 [astro-ph.CO]].


\bibitem{Han:2018yrk}
C.~Han, S.~Pi and M.~Sasaki,
Phys. Lett. B \textbf{791} (2019), 314-318
doi:10.1016/j.physletb.2019.02.037 [arXiv:1809.05507 [hep-ph]].



\bibitem{Achucarro:2018vey}
A.~Ach\'ucarro and G.~A.~Palma,
JCAP \textbf{02} (2019), 041 doi:10.1088/1475-7516/2019/02/041
[arXiv:1807.04390 [hep-th]].



\bibitem{Akrami:2020zfz}
Y.~Akrami, M.~Sasaki, A.~R.~Solomon and V.~Vardanyan,
[arXiv:2008.13660 [astro-ph.CO]].



\bibitem{Colgain:2018wgk}
E.~\'O Colg\'ain, M.~H.~P.~M.~van Putten and H.~Yavartanoo,
Phys. Lett. B \textbf{793} (2019), 126-129
doi:10.1016/j.physletb.2019.04.032 [arXiv:1807.07451 [hep-th]].



\bibitem{Colgain:2019joh}
E.~\'O.~Colg\'ain and H.~Yavartanoo,
Phys. Lett. B \textbf{797} (2019), 134907
doi:10.1016/j.physletb.2019.134907 [arXiv:1905.02555
[astro-ph.CO]].


\bibitem{Banerjee:2020xcn}
A.~Banerjee, H.~Cai, L.~Heisenberg, E.~\'O.~Colg\'ain,
M.~M.~Sheikh-Jabbari and T.~Yang,
[arXiv:2006.00244 [astro-ph.CO]].



\bibitem{Oikonomou:2021zfl}
V.~K.~Oikonomou, I.~Giannakoudi, A.~Gitsis and K.~R.~Revis,
Int. J. Mod. Phys. D \textbf{31} (2022) no.02, 2250001
doi:10.1142/S0218271822500018 [arXiv:2105.11935 [gr-qc]].

\bibitem{Gitsis:2023cyw}
A.~Gitsis, K.~R.~Revis, S.~A.~Venikoudis and F.~P.~Fronimos,
[arXiv:2301.08126 [gr-qc]].























\bibitem{Venikoudis:2022gfg}
S.~A.~Venikoudis and F.~P.~Fronimos,
Nucl. Phys. B \textbf{984} (2022), 115945
doi:10.1016/j.nuclphysb.2022.115945 [arXiv:2208.08333 [gr-qc]].





\bibitem{ElBourakadi:2021blc}
K.~El Bourakadi,
[arXiv:2104.10552 [gr-qc]].

\bibitem{ElBourakadi:2021nyb}
K.~El Bourakadi, M.~Ferricha-Alami, H.~Filali, Z.~Sakhi and M.~Bennai,
Eur. Phys. J. C \textbf{81} (2021) no.12, 1144
doi:10.1140/epjc/s10052-021-09946-4
[arXiv:2209.08581 [gr-qc]].

\bibitem{Koh:2018qcy}
S.~Koh, B.~H.~Lee and G.~Tumurtushaa,
Phys. Rev. D \textbf{98} (2018) no.10, 103511
doi:10.1103/PhysRevD.98.103511
[arXiv:1807.04424 [astro-ph.CO]].

\bibitem{Cook:2015vqa}
J.~L.~Cook, E.~Dimastrogiovanni, D.~A.~Easson and L.~M.~Krauss,
JCAP \textbf{04} (2015), 047
doi:10.1088/1475-7516/2015/04/047
[arXiv:1502.04673 [astro-ph.CO]].


\bibitem{ElBourakadi:2022lqf}
K.~El Bourakadi, Z.~Sakhi and M.~Bennai,
Int. J. Mod. Phys. A \textbf{37} (2022) no.17, 2250117
doi:10.1142/S0217751X22501172
[arXiv:2209.09241 [gr-qc]].


\bibitem{Mazumdar:2013gya}
A.~Mazumdar and B.~Zaldivar,
Nucl. Phys. B \textbf{886} (2014), 312-327
doi:10.1016/j.nuclphysb.2014.07.001
[arXiv:1310.5143 [hep-ph]].


\bibitem{Ellis:2015pla}
J.~Ellis, M.~A.~G.~Garcia, D.~V.~Nanopoulos and K.~A.~Olive,
JCAP \textbf{07} (2015), 050
doi:10.1088/1475-7516/2015/07/050
[arXiv:1505.06986 [hep-ph]].

\bibitem{Hasegawa:2019jsa}
T.~Hasegawa, N.~Hiroshima, K.~Kohri, R.~S.~L.~Hansen, T.~Tram and S.~Hannestad,
JCAP \textbf{12} (2019), 012
doi:10.1088/1475-7516/2019/12/012
[arXiv:1908.10189 [hep-ph]].
























\bibitem{Odintsov:2021kup}
S.~D.~Odintsov, V.~K.~Oikonomou and F.~P.~Fronimos,
Phys. Dark Univ. \textbf{35} (2022), 100950
doi:10.1016/j.dark.2022.100950 [arXiv:2108.11231 [gr-qc]].




\bibitem{Kamionkowski:2015yta}
M.~Kamionkowski and E.~D.~Kovetz,
Ann. Rev. Astron. Astrophys. \textbf{54} (2016), 227-269
doi:10.1146/annurev-astro-081915-023433 [arXiv:1510.06042
[astro-ph.CO]].




\bibitem{Turner:1993vb}
M.~S.~Turner, M.~J.~White and J.~E.~Lidsey,
Phys. Rev. D \textbf{48} (1993), 4613-4622
doi:10.1103/PhysRevD.48.4613 [arXiv:astro-ph/9306029 [astro-ph]].

\bibitem{Boyle:2005se}
L.~A.~Boyle and P.~J.~Steinhardt,
Phys. Rev. D \textbf{77} (2008), 063504
doi:10.1103/PhysRevD.77.063504 [arXiv:astro-ph/0512014
[astro-ph]].



\bibitem{Zhang:2005nw}
Y.~Zhang, Y.~Yuan, W.~Zhao and Y.~T.~Chen,
Class. Quant. Grav. \textbf{22} (2005), 1383-1394
doi:10.1088/0264-9381/22/7/011 [arXiv:astro-ph/0501329
[astro-ph]].



\bibitem{Caprini:2018mtu}
C.~Caprini and D.~G.~Figueroa,
Class. Quant. Grav. \textbf{35} (2018) no.16, 163001
doi:10.1088/1361-6382/aac608 [arXiv:1801.04268 [astro-ph.CO]].




\bibitem{Clarke:2020bil}
T.~J.~Clarke, E.~J.~Copeland and A.~Moss,
JCAP \textbf{10} (2020), 002 doi:10.1088/1475-7516/2020/10/002
[arXiv:2004.11396 [astro-ph.CO]].



\bibitem{Smith:2005mm}
T.~L.~Smith, M.~Kamionkowski and A.~Cooray,
Phys. Rev. D \textbf{73} (2006), 023504
doi:10.1103/PhysRevD.73.023504 [arXiv:astro-ph/0506422
[astro-ph]].




\bibitem{Giovannini:2008tm}
M.~Giovannini,
Class. Quant. Grav. \textbf{26} (2009), 045004
doi:10.1088/0264-9381/26/4/045004 [arXiv:0807.4317 [astro-ph]].

\bibitem{Liu:2015psa}
X.~J.~Liu, W.~Zhao, Y.~Zhang and Z.~H.~Zhu,
Phys. Rev. D \textbf{93} (2016) no.2, 024031
doi:10.1103/PhysRevD.93.024031 [arXiv:1509.03524 [astro-ph.CO]].




\bibitem{Vagnozzi:2020gtf}
S.~Vagnozzi,
Mon. Not. Roy. Astron. Soc. \textbf{502} (2021) no.1, L11-L15
doi:10.1093/mnrasl/slaa203 [arXiv:2009.13432 [astro-ph.CO]].





\bibitem{Kamionkowski:1993fg}
M.~Kamionkowski, A.~Kosowsky and M.~S.~Turner,
Phys. Rev. D \textbf{49} (1994), 2837-2851
doi:10.1103/PhysRevD.49.2837 [arXiv:astro-ph/9310044 [astro-ph]].

\bibitem{Giare:2020vss}
W.~Giar\`e and F.~Renzi,
Phys. Rev. D \textbf{102} (2020) no.8, 083530
doi:10.1103/PhysRevD.102.083530 [arXiv:2007.04256 [astro-ph.CO]].


\bibitem{Zhao:2006mm}
W.~Zhao and Y.~Zhang,
Phys. Rev. D \textbf{74} (2006), 043503
doi:10.1103/PhysRevD.74.043503 [arXiv:astro-ph/0604458
[astro-ph]].






\bibitem{Lasky:2015lej}
P.~D.~Lasky, C.~M.~F.~Mingarelli, T.~L.~Smith, J.~T.~Giblin,
D.~J.~Reardon, R.~Caldwell, M.~Bailes, N.~D.~R.~Bhat,
S.~Burke-Spolaor and W.~Coles, \textit{et al.}
Phys. Rev. X \textbf{6} (2016) no.1, 011035
doi:10.1103/PhysRevX.6.011035 [arXiv:1511.05994 [astro-ph.CO]].





\bibitem{Cai:2021uup}
R.~G.~Cai, C.~Fu and W.~W.~Yu,
[arXiv:2112.04794 [astro-ph.CO]].



\bibitem{Benetti:2021uea}
M.~Benetti, L.~L.~Graef and S.~Vagnozzi,
Phys. Rev. D \textbf{105} (2022) no.4, 043520
doi:10.1103/PhysRevD.105.043520 [arXiv:2111.04758 [astro-ph.CO]].





\bibitem{Lin:2021vwc}
J.~Lin, S.~Gao, Y.~Gong, Y.~Lu, Z.~Wang and F.~Zhang,
[arXiv:2111.01362 [gr-qc]].

\bibitem{Zhang:2021vak}
F.~Zhang, J.~Lin and Y.~Lu,
Phys. Rev. D \textbf{104} (2021) no.6, 063515 [erratum: Phys. Rev.
D \textbf{104} (2021) no.12, 129902]
doi:10.1103/PhysRevD.104.063515 [arXiv:2106.10792 [gr-qc]].




\bibitem{Pritchard:2004qp}
J.~R.~Pritchard and M.~Kamionkowski,
Annals Phys. \textbf{318} (2005), 2-36
doi:10.1016/j.aop.2005.03.005 [arXiv:astro-ph/0412581 [astro-ph]].

\bibitem{Khoze:2022nyt}
V.~V.~Khoze and D.~L.~Milne,
[arXiv:2212.04784 [hep-ph]].








\bibitem{ElBourakadi:2022anr}
K.~El Bourakadi, B.~Asfour, Z.~Sakhi, Z.~M.~Bennai and T.~Ouali,
Eur. Phys. J. C \textbf{82} (2022) no.9, 792
doi:10.1140/epjc/s10052-022-10762-7 [arXiv:2209.08585 [gr-qc]].



\bibitem{Arapoglu:2022vbf}
A.~S.~Arapo\u{g}lu and A.~E.~Y\"ukselci,
[arXiv:2210.16699 [gr-qc]].


\bibitem{Giare:2022wxq}
W.~Giar\`e, M.~Forconi, E.~Di Valentino and A.~Melchiorri,
[arXiv:2210.14159 [astro-ph.CO]].




\bibitem{Odintsov:2022cbm}
S.~D.~Odintsov, V.~K.~Oikonomou and R.~Myrzakulov,
Symmetry \textbf{14} (2022) no.4, 729 doi:10.3390/sym14040729
[arXiv:2204.00876 [gr-qc]].



\bibitem{Linder:2021pek}
E.~V.~Linder,
[arXiv:2108.11526 [astro-ph.CO]].


\bibitem{reviews1}
 S.~Nojiri, S.~D.~Odintsov and V.~K.~Oikonomou,
  Phys.\ Rept.\  {\bf 692} (2017) 1
  [arXiv:1705.11098 [gr-qc]].

\bibitem{reviews2}


 S. Capozziello, M. De Laurentis,
   Phys.\ Rept.\  {\bf 509}, 167 (2011);\\




\bibitem{reviews3}
 V.~Faraoni and S.~Capozziello,
  Fundam.\ Theor.\ Phys.\  {\bf 170} (2010).


   \bibitem{reviews4}

S. Nojiri, S.D. Odintsov,
   Phys.\ Rept.\  {\bf 505}, 59 (2011);







\bibitem{Oikonomou:2020oex}
V.~K.~Oikonomou,
Phys. Rev. D \textbf{103} (2021) no.12, 124028
doi:10.1103/PhysRevD.103.124028 [arXiv:2012.01312 [gr-qc]].





\bibitem{Codello:2015mba}
A.~Codello and R.~K.~Jain,
Class. Quant. Grav. \textbf{33} (2016) no.22, 225006
doi:10.1088/0264-9381/33/22/225006 [arXiv:1507.06308 [gr-qc]].

\bibitem{Horndeski:1974wa}
G.~W.~Horndeski,
Int. J. Theor. Phys. \textbf{10} (1974), 363-384
doi:10.1007/BF01807638

\bibitem{Odintsov:2021nim}
S.~D.~Odintsov, V.~K.~Oikonomou and F.~P.~Fronimos,
Class. Quant. Grav. \textbf{38} (2021) no.7, 075009
doi:10.1088/1361-6382/abe24f [arXiv:2102.02239 [gr-qc]].



\bibitem{Oikonomou:2022tux}
V.~K.~Oikonomou,
Phys. Rev. D \textbf{106} (2022) no.4, 044041
doi:10.1103/PhysRevD.106.044041 [arXiv:2208.05544 [gr-qc]].



\bibitem{DeFelice:2011zh}
A.~De Felice and S.~Tsujikawa,
JCAP \textbf{04} (2011), 029 doi:10.1088/1475-7516/2011/04/029
[arXiv:1103.1172 [astro-ph.CO]].




\bibitem{Planck:2018vyg}
N.~Aghanim \textit{et al.} [Planck],
Astron. Astrophys. \textbf{641} (2020), A6 [erratum: Astron.
Astrophys. \textbf{652} (2021), C4]
doi:10.1051/0004-6361/201833910 [arXiv:1807.06209 [astro-ph.CO]].








\bibitem{Zarei:2014bta}
M.~Zarei,
Class. Quant. Grav. \textbf{33} (2016) no.11, 115008
doi:10.1088/0264-9381/33/11/115008 [arXiv:1408.6467
[astro-ph.CO]].






\bibitem{Lyth:1996im}
D.~H.~Lyth,
Phys. Rev. Lett. \textbf{78} (1997), 1861-1863
doi:10.1103/PhysRevLett.78.1861 [arXiv:hep-ph/9606387 [hep-ph]].




\bibitem{Steinhardt:1999nw}
P.~J.~Steinhardt, L.~M.~Wang and I.~Zlatev,
Phys. Rev. D \textbf{59} (1999), 123504
doi:10.1103/PhysRevD.59.123504 [arXiv:astro-ph/9812313
[astro-ph]].


\bibitem{Oikonomou:2021yks}
V.~K.~Oikonomou,
Int. J. Geom. Meth. Mod. Phys. \textbf{19} (2022) no.07, 2250099
doi:10.1142/S0219887822500992 [arXiv:2106.10778 [gr-qc]].






\end{thebibliography}
\end{document}